\documentclass[12pt]{article}

\usepackage{amsfonts,amssymb,amsmath}
\usepackage[dvips]{epsfig}
\begin{document}
\title{\bf Quantum coherence and entanglement preservation in Markovian and non-Markovian dynamics via additional qubits}
\author{ N. Behzadi $^{a}$
\thanks{E-mail:n.behzadi@tabrizu.ac.ir}  ,
B. Ahansaz $^{b}$ and
E. Faizi $^{b}$
\\ $^a${\small Research Institute for Fundamental Sciences, University of Tabriz, Tabriz, Iran,}
\\ $^b${\small Physics Department, Azarbaijan Shahid Madani University, Tabriz, Iran}} \maketitle

\begin{abstract}
\noindent
In this paper, we investigate preservation of quantum coherence of a single-qubit interacting with a zero-temperature thermal reservoir through the addition of non-interacting qubits in the reservoir. Moreover, we extend this scheme to preserve quantum entanglement between two and three distant qubits, each of which interacts with a dissipative reservoir independently. At the limit $t\rightarrow\infty$, we obtained analytical expressions for the coherence measure and the concurrence of two and three qubits in terms of the number of additional qubits.
It is observed that, by increasing the number of additional qubits in each reservoir, the
initial coherence and the respective entanglements are completely protected in both Markovian
and non-Markovian regimes. Interestingly, the protection of entanglements occurs even under the individually different behaviors of the reservoirs.
\\
\\
{\bf PACS Nos:}
\\
{\bf Keywords:} Quantum coherence preservation, Entanglement preservation, Measure of coherence, Concurrence, Lower bound of concurrence, Thermal reservoir, Additional qubits, Markovian, non-Markovian.
\end{abstract}

\section{Introduction}
Quantum technology relies on the utilization of resources, like quantum coherence and entanglement, which improve considerably quantum information processing protocols relative classical ones $\cite{Perrin, Nielsen, Benenti}$.
However, the quantum entanglement is so fragile and undergoes either an asymptotic decay or a sudden death $\cite{Hornberger, Merali}$.
This is due to decoherence, whereby the unavoidable interaction between any real quantum system with its surrounding environment alters the quantum system and consequently disentanglement occurs. Therefore, it is very important to investigate the physical systems and the physical effects that
may lead to effective long-time entanglement preservation.
Although there are so-called decoherence-free states which the initial entanglement remains invariant in time, however, there is only a certain kind of entangled state which can be
decoupled from the influence of the environment in this way $\cite{Maniscalco, Li, An1}$.
So far, a lot of researches have been devoted to entanglement manipulation and protection.
For instance, the quantum Zeno effect (QZE) is a promising way to avoid the decaying behaviour of the entanglement in dissipative systems. This effect which refers to the inhibition of the temporal evolution of a quantum system by repeated projective measurements during a defined period of time, has been discussed in $\cite{Maniscalco, Mundarain, Rossi, Hou}$.
But, this scheme is relatively difficult since one has to perform a series of measurements to the system during the course of the evolution.
The other scheme for protecting the entanglement is detuning modulation $\cite{An1, Xiao}$, where the quantum entanglement is not preserved in the long-time limit.
Another proposed scheme, without the disadvantages of the
aforementioned schemes, focuses on the long time limit protection of quantum
entanglement through the additional qubits. The protection of entanglement between two qubits via the additional qubits was first observed in $\cite{An2}$, where the bipartite entanglement is sustained by the addition of a third qubit. Improving the preservation of entanglement by much more additional qubits introduced
in $\cite{An3, Flores}$, where all of the entangled qubits and the additional ones were contained in a common environment.

Distantly non-interacting two-level quantum systems, each of which interacts with an
environment independently, are preferable elements for a quantum hardware in order to
accomplish the individual control required for quantum information processing $\cite{Ladd, Xiang}$.
Therefore, finding strategies to protect quantum resources, such as coherence and entanglement, is an essential task in these configurations $\cite{Franco}$.

In this paper, we consider a system of $N$ non-interacting qubits immersed in a common zero-temperature thermal Lorentzian reservoir and show how quantum coherence of a given single-qubit in this system can be preserved by increasing the number of additional $N-1$ qubits. At the long time limit ($t\rightarrow\infty$), we show that, apart from Markovian and non-Markovian dynamics, the coherence measure for the single-qubit approaches asymptotically to a non-zero steady value depending on the number of additional qubits and the initial state coherence of the single-qubit.
Also, by considering $N\rightarrow\infty$, our calculations show that the coherence measure reaches to its initial value.

In the next step, we extend this scheme to protect bipartite and tripartite entanglements
among two and three distant qubits, each of which contained in an independent reservoir.
It is observed that adding other qubits to any of the reservoirs, leads to protecting entanglement from sudden death. At the limit $t\rightarrow\infty$, we obtain simple analytical expressions for each of the bipartite concurrence $\cite{Wootterrs}$ and tripartite lower bound of concurrence (LBC) $\cite{gao}$
in terms of the number of additional qubits contained in each of the reservoirs and
the respective initial state entanglement. Moreover, increasing the number of additional qubits in
each of the reservoirs completely preserves the initial state entanglement from the dissipative processes
of the reservoirs.
As illustrated in the text, this mechanism of entanglement preservation works even for different Markovian and non-Markovian dynamics occurred individually in each of the related subsystems. After this Introduction, in Sec. II we formulate our scheme for preserving the quantum coherence of a single-qubit. Sec. III is devoted for extending the scheme proposed in Sec II for protection of bipartite and tripartite entanglements. And finally, the paper is ended with a brief conclusion in Sec. IV.

\section{Quantum coherence preservation}
In this section, we consider a system consisting of $N$ independent qubits (two-level atoms) immersed in a common zero-temperature thermal reservoir,
as depicted in Fig. 1(a). The Hamiltonian $\hat{H}$ of the system contains two parts ($\hbar=1$)
\begin{eqnarray}
\hat{H}=\hat{H}_{0}+\hat{H}_{I},
\end{eqnarray}
with $\hat{H}_{0}$ the free Hamiltonian and $\hat{H}_{I}$ describes the interaction Hamiltonian
\begin{eqnarray}
\hat{H}_{0}=\Omega \sum_{j=1}^{N} {\hat{\sigma}_{j}^{+} \hat{\sigma}_{j}^{-}}+\sum_{k} \omega_{k} \hat{b_{k}}^{\dagger} \hat{b_{k}},
\end{eqnarray}

\begin{eqnarray}
\hat{H}_{I}= \sum_{j=1}^{N} \sum_{k} \beta_{j}(g_{k} {\hat{\sigma}_{j}^{+}} \hat{b_{k}}+g_{k}^{*} {\hat{\sigma}_{j}^{-}} \hat{b_{k}}^{\dagger}),
\end{eqnarray}
where $\hat{\sigma}_{j}^{+}(\hat{\sigma}_{j}^{-})$ is the raising (lowering) operator of the $j^{th}$ qubit with transition frequency $\Omega$
and $\hat{b_{k}}$ ($\hat{b_{k}^{\dagger}}$) is the annihilation (creation) operator of the $k^{th}$ field mode with frequency $\omega_{k}$.
The strength of coupling between the $j^{th}$ qubit and the $k^{th}$ field mode is described by $g_{k}$ and
the dimensionless real constants $\beta_{j}$ are introduced to individualize the qubits.

It is convenient to work in the interaction picture where the state $|\psi(t)\rangle$ of the system obeys the Schrodinger equation as
\begin{eqnarray}
  i \frac{d}{dt}|\psi(t)\rangle=\hat{H}_{I}(t) |\psi(t)\rangle,
\end{eqnarray}
and the Hamiltonian in this picture is given by
\begin{eqnarray}
\hat{H}_{I}(t)=e^{i \hat{H}_{0}t} \hat{H}_{I} e^{-i \hat{H}_{0}t}=\sum_{j=1}^{N} \sum_{k} \beta_{j}(g_{k} {\hat{\sigma}_{j}^{+}} \hat{b_{k}} e^{i(\Omega-\omega_{k})t}+g_{k}^{*} {\hat{\sigma}_{j}^{-}} \hat{b_{k}}^{\dagger} e^{-i(\Omega-\omega_{k})t}).
\end{eqnarray}
Since the total Hamiltonian commutes with the number of excitations, i.e. $\big[(\sum_{j=1}^{N} {\hat{\sigma}_{j}^{+} \hat{\sigma}_{j}^{-}}+\sum_{k} \hat{b_{k}}^{\dagger} \hat{b_{k}}),H\big]=0$,
therefore, any initial state of the form
\begin{eqnarray}
  |\psi(0)\rangle=C_{0}(0)|0\rangle_{S} |0\rangle_{E}+\sum_{j=1}^{N}C_{j}(0)|j\rangle_{S} |0\rangle_{E},
\end{eqnarray}
evolves after time $t$ into the following state
\begin{eqnarray}
\begin{array}{c}
  |\psi(t)\rangle=C_{0}(t)|0\rangle_{S} |0\rangle_{E}+\sum_{j=1}^{N}C_{j}(t)|j\rangle_{S} |0\rangle_{E}+\sum_{k} D_{k}(t) |0\rangle_{S} |1_{k}\rangle_{E},
\end{array}
\end{eqnarray}
where $|j\rangle_{S}=|g\rangle^{\bigotimes N}_{j^{th}\equiv e}$, which means that all of the qubits are in the ground state $|g\rangle$ except the $j^{th}$ qubit,
which is in the excited state $|e\rangle$ and $|0\rangle_{S}=|g\rangle^{\bigotimes N}=|g,g,...,g\rangle$.
Also, we denote $|0\rangle_{E}$ being the vacuum state of the reservoir and $|1_{k}\rangle_{E}$ the state of it with only one excitation in the $k^{th}$ field mode.

Substituting Eqs. (5) and (7) into Eq. (4), gives the following set of $N+1$ differential equations for the probability amplitudes as
\begin{eqnarray}
  \dot{C}_{j}(t)=-i \beta_{j} \sum_{k} g_{k} D_{k}(t) e^{i(\Omega-\omega_{k})t},\\
  \dot{D}_{k}(t)=-i \sum_{j=1}^{N} \beta_{j} g_{k}^{*} C_{j}(t) e^{-i(\Omega-\omega_{k})t},
\end{eqnarray}
where $j=1,2,...,N$. It is clear that $\dot{C}_{0}(t)=0$, then $C_{0}(t)=C_{0}(0)=C_{0}$. Integrating Eq. (9) and substituting it into the Eq. (8), gives
the following set of $N$ integro-differential equations
\begin{eqnarray}
  \frac{dC_{j}(t)}{dt}=-\int_{0}^{t} f(t-t')\beta_{j} \sum_{l=1}^{N} \beta_{l} C_{l}(t') dt',
\end{eqnarray}
and the correlation function $f(t-t')$ is related to the spectral density $J(\omega)$ of the reservoir by
\begin{eqnarray}
f(t-t')=\int_{-\infty}^{\infty} d\omega J(\omega) e^{i(\Omega-\omega)(t-t')}.
\end{eqnarray}
The exact form of $C_{j}(t)$ thus depends on the particular choice for the spectral density of the reservoir. We take the Lorentzian spectral density for the reservoir as
\begin{eqnarray}
  J(\omega)=\frac{1}{2\pi} \frac{\gamma_{0} \lambda^{2}}{(\omega-\Omega+\Delta)^2+\lambda^{2}},
\end{eqnarray}
where $\Delta$ is the detuning between the transition frequency of the qubits $\Omega$ and the central frequency of the reservoir.
The parameter $\lambda$ defines the spectral width of the coupling and the parameter $\gamma$ is the coupling constant.
By using the spectral density $J(\omega)$ given by Eq. (12), the exact solutions of the probability amplitudes $C_{j}(t)$ can be obtained (see appendix A) as
\begin{eqnarray}
\begin{array}{c}
  C_{j}(t)=e^{-\Lambda t/2}\bigg(\mathrm{cosh}{(\frac{Dt}{2})}+\frac{\Lambda}{D} \mathrm{sinh}{(\frac{Dt}{2}})\bigg) C_{j}(0)+\\\\
  \frac{\sum_{l\neq j}^{N} {\beta_{l}}^2 C_{j}(0)-\beta_{l} \beta_{j} C_{l}(0)}{\sum_{l=1}^{N} {\beta_{l}}^2}\bigg(1-e^{-\Lambda t/2}\Big(\mathrm{cosh}{(\frac{Dt}{2})}+\frac{\Lambda}{D} \mathrm{sinh}{(\frac{Dt}{2})}\Big)\bigg),
\end{array}
\end{eqnarray}
where $\Lambda=\lambda-i\Delta$ and $D=\sqrt{\Lambda^{2}-2\gamma_{0} \lambda \sum_{l=1}^{N} {\beta_{l}}^2}$.
Here, there are two regimes for the system environment coupling $\cite{An1,An2,Flores}$: weak coupling regime ($\lambda>2\gamma_{0} \sum_{l=1}^{N} {\beta_{l}}^2$)
and strong coupling regime ($\lambda<2\gamma_{0} \sum_{l=1}^{N} {\beta_{l}}^2$).
In the weak coupling regime the behaviour of the system is Markovian and irreversible decay occurs. However, in
the strong coupling regime the non-Markovian dynamics occurs accompanied by an oscillatory reversible decay.
After tracing out the zero-temperature thermal reservoir and the qubits except $j^{th}$ one, the reduced density
matrix of the $j^{th}$ qubit in the basis $\{|e\rangle, |g\rangle \}$, at time $t$, becomes
\begin{eqnarray}
\rho_{j}(t)=\left(
                \begin{array}{cc}
                  |C_{j}(t)|^2 & C_{0}^{*} C_{j}(t) \\\\
                  C_{0} C_{j}^{*}(t) & 1-|C_{j}(t)|^2 \\
                \end{array}
              \right).
\end{eqnarray}
Furthermore, if we let $\beta_{l}=1$ for $l=1,2,...,N$ and $C_{l}(0)=0$ with $l\neq j$, Eq. (13) reduces to
\begin{eqnarray}
  C_{j}(t)=G(t)C_{j}(0),
\end{eqnarray}
where
\begin{eqnarray}
  G(t)=\frac{N-1}{N}+\frac{e^{-\Lambda t/2}}{N} \Big(\mathrm{cosh}{(\frac{Dt}{2})}+\frac{\Lambda}{D} \mathrm{sinh}{(\frac{Dt}{2})}\Big),
\end{eqnarray}
and $D=\sqrt{\Lambda^{2}-2\gamma_{0} \lambda N}$. The reduced density matrix of the $j^{th}$ qubit in Eq. (14), can be rewritten as
\begin{eqnarray}
\rho_{j}(t)=\left(
                \begin{array}{cc}
                  |G(t)|^{2}|C_{j}(0)|^2 &  C_{0}^{*}G(t)C_{j}(0) \\\\
                   C_{0}G(t)^{\ast}C_{j}^{*}(0) & 1-|G(t)|^{2}|C_{j}(0)|^2 \\
                \end{array}
              \right).
\end{eqnarray}
The $j^{th}$ qubit dynamics thus depends only on the function ${G(t)}$ ($0 < |G(t)| \leq 1$), which in turns depends on the spectral density and the number of qubits. Also, as discussed in $\cite{Breuer}$, by differentiating of Eq. (17) with respect to time and comparing it with
an exact master equation, decay rate of the $j^{th}$ qubit can be obtained as
\begin{eqnarray}
\Gamma(t)=-2 Re\Big\{\frac{\dot{C}_{j}(t)}{C_{j}(t)}\Big\}=Re\bigg\{\frac{2\gamma_{0} \lambda e^{-\Lambda t/2} sinh{(\frac{Dt}{2})}}{D\bigg(\frac{N-1}{N}+\frac{e^{-\Lambda t/2}}{N} \Big(\mathrm{cosh}{(\frac{Dt}{2})}+\frac{\Lambda}{D} \mathrm{sinh}{(\frac{Dt}{2}})\Big)\bigg)}\bigg\}.
\end{eqnarray}

Baumgratz $et.al$ introduced an intuitive measure of quantum coherence based on the off-diagonal elements of density matrix for the desired quantum state $\cite{Baumgratz}$, as
\begin{eqnarray}
\xi(\rho(t))=\sum_{m,n(m\neq n)} |\rho_{m,n}(t)|,
\end{eqnarray}
where $\rho_{m,n}(t) (m\neq n)$ are the off-diagonal elements of the system density matrix. Indeed, it has been
recently shown that $\xi(\rho(t))$ satisfies the physical requirements which make it as a proper coherence measure.

Suppose the $j^{th}$ qubit is initially prepared in the state $\alpha_{1} |g\rangle+\alpha_{2} |e\rangle$ ($|\alpha_{1}|^2+|\alpha_{2}|^2=1$), the other qubits are prepared in the state $|g\rangle$ and the reservoir also is in its respective vacuum state, so in this way, $\xi(\rho_{j}(0))=2|C_{0} C_{j}(0)|=2|\alpha_{1} \alpha_{2}|$. Therefore, at time $t>0$, the coherence of the $j^{th}$ qubit becomes $\xi(\rho_{j}(t))=2|C_{0} C_{j}(t)|=2|G(t)||C_{0} C_{j}(0)|$.
The dynamical behavior of the coherence in terms of the dimensionless time $\gamma_{0} t$
has been shown in Fig. 2, where we have assumed $C_{0}=C_{j}(0)=1/\sqrt{2}$.
In Fig. 2 (a, b), the coherence measure of the qubit exhibits a Markovian dynamics and monotonically approaches to zero in the absence of additional qubits ($N=1$).
However, it is readily observed that it can be greatly inhibited by increasing the number of additional qubits ($N=2,3,6$).
Moreover, in the non-Markovian regime and in the absence of additional qubits ($N=1$), the measure of coherence oscillatory undergoes to sudden death and in the presence of additional qubits ($N=2,3,6$), undergoes to a non-zero steady value, as depicted in Fig. 2(c, d).
The oscillations of coherence in the non-Markovian regime constitute a sufficient condition to signify the presence of memory effects in the system dynamics,
being due to information backflow from the environment to the system.
We point out that the existence of detuning ($\Delta$) tend to slow down the decay process of the coherence in the non-Markovian regime.
It should be noted that, in the long time limit, the measure of coherence reduces to the following expression
\begin{eqnarray}
  \xi(\rho_{j}(t))=2|C_{0} C_{j}(t)|=2|\alpha_{1} \alpha_{2}|(\frac{N-1}{N}).
\end{eqnarray}
Indeed, the single-qubit coherency can be determined only by knowing the number of additional $N-1$ qubits in the reservoir and the initial state of the qubit.
Also, as $N\rightarrow \infty$, $\xi(\rho_{j}(t))$ reaches to $2|\alpha_{1} \alpha_{2}|$ (i.e. the initial coherency of the $j^{th}$ qubit).
On the other hand, these observations are confirmed by the behavior of the related decay rate (18) in terms of the scaled parameter $\gamma_{0} t$, as shown in Fig. 3.

Before extending the obtained results for protecting entanglement in the next sections, let us make a discussion in order to clarify the connection between the proposed scheme in this paper and the decoherence-free subspace method $\cite{lid1, lid2, lid3}$. According to these references, consider that the related Hilbert space of the qubit system regarded as our open system is as $\mathcal{H}_{S}=\mathcal{H}_{D-F}\oplus \mathcal{H}_{N}$ where $\mathcal{H}_{D-F}$ is the decoherence-free subspace and $\mathcal{H}_{N}$ denotes the noisy one. When we have only the considered single-qubit, the initial state of this qubit completely belongs to the noisy subspace $\mathcal{H}_{N}$. Therefore, at long time limit and according to Eq. (20), the qubit ultimately loses its coherency as shown in Fig. 2. In fact, in this stage, there is no decoherence-free subspace $\mathcal{H}_{D-F}$. As the non-interacting additional qubits enter to the reservoir some decoherence-free or subradiant states are created and we have a $\mathcal{H}_{D-F}$ subspace where the initial state of the single-qubit has a non-zero support in that subspace. So, in this way, it is observed a non-zero steady value for the coherence measure (see Eq. (20) and Fig. 2). Consequently, when the number of additional qubits becomes very large (infinity), the $\mathcal{H}_{D-F}$ subspace will be sufficiently large such that the initial state of the respective single-qubit completely belongs to this subspace so its coherence remains unchanged. It is  concluded, in this regard, that the decoherence-free subspace can be effectively manipulated through the additional qubits. This argument is also valid for entanglement protection which will be discussed in the next sections.

Since the appearance of decoherence-free subspace $\mathcal{H}_{D-F}$ depends only on the presence of additional qubits so it is expected that it can not be depend on the structure of the reservoir as confirmed in $\cite{bh}$. Therefore, the preservation of quantum coherence and entanglement, in long time limit, can not be dependent on the spectral shape of the reservoir such as Lorentzian, Ohmic, sub-Ohmic or super-Ohmic cases.

\section{Quantum entanglement preservation}
\subsection{EPR-type entanglement}
To achieve to the scheme of entanglement preservation, we consider a composite system consisting of two subsystems $A$ and $B$ contained in two independent Lorenzian reservoirs.
Each of the subsystems contains $N_{A}$ and $N_{B}$ qubits, respectively.
Let's consider the $j^{th}$ qubit of subsystem $A$ and the $l^{th}$ qubit of subsystem $B$ prepared initially in an EPR-type entangled state as follows
\begin{eqnarray}
  |\phi(0)\rangle_{j,l}=C_{j}^{A}(0)|e,g\rangle+C_{l}^{B}(0)|g,e\rangle,
\end{eqnarray}
as depicted in Fig. 1 (b).
According to the Ref. $\cite{Bellomo}$, the complete dynamics of the above two-qubit system can be obtained easily  (see appendix B). Therefore, in the standard computational basis such as $\{|1\rangle \equiv|e,e\rangle, |2\rangle \equiv |e,g\rangle, |3\rangle \equiv |g,e\rangle, |4\rangle \equiv |g,g\rangle\}$, the explicit forms of the matrix elements of the density operator at time $t$ becomes
\begin{eqnarray}
\begin{array}{c}
\rho_{22}(t)=|G^{A}(t)|^{2}|C_{j}^{A}(0)|^2,\\\\
\rho_{33}(t)=|G^{B}(t)|^{2}|C_{l}^{B}(0)|^2,\\\\
\rho_{44}(t)=1-|G^{A}(t)|^{2}|C_{j}^{A}(0)|^2-|G^{B}(t)|^{2}|C_{l}^{B}(0)|^2,\\\\
\rho_{23}(t)=\rho_{32}^{*}(t)=G^{A}(t) {G^{B}}^{*}(t) C_{j}^{A}(0) {C_{l}^{B}}^{*}(0),\\\\
\rho_{11}(t)=\rho_{12}(t)=\rho_{13}(t)=\rho_{14}(t)=\rho_{24}(t)=\rho_{34}(t)=0.
\end{array}
\end{eqnarray}
where $G^{A}(t)$, $G^{B}(t)$ are determined in similar way as case of the Eq. (16).
The parameter $\lambda_{A} (\lambda_{B})$ which appears in the $G^{A}(t)$ ($G^{B}(t)$) defines the spectral width for the coupling of the qubits at subsystem $A$ (subsystem $B$) to its respective reservoir. Also, we assume that the coupling constants of the qubits to their respective reservoirs are equal, i.e. $\gamma_{A}=\gamma_{B}=\gamma_{0}$.

To quantify the amount of entanglement for the state (22), we use concurrence as a measure of two-qubit entanglement. The following analytical form for the concurrence is obtained
\begin{eqnarray}
  C(\rho_{j,l}(t))=2 |G^{A}(t) G^{B}(t)| |C_{j}^{A}(0) C_{l}^{B}(0)|.
\end{eqnarray}
The time dependency of the concurrence has been shown in Fig. 4 in the Markovian and Non-Markovian regimes.
Notice that in the case of $N_{A}=N_{B}=1$ (in the absence of additional qubits in the subsystems A and B), the concurrence ultimately decays to zero but there is a steady non-zero value for it by increasing $N_{A}$ and $N_{B}$ in both of the Markovian and non-Markovian regimes.
An interesting result for the concurrence occurs at the asymptotical limit $t\rightarrow \infty$, as follows
\begin{eqnarray}
  C(\rho_{j,l})=2 \frac{(N_{A}-1)(N_{B}-1)}{N_{A}N_{B}} |C_{j}^{A}(0) C_{l}^{B}(0)|.
\end{eqnarray}
Indeed, in the long time limit, the concurrence can be determined only by knowing the number of qubits in the reservoirs and the initial state of the two-qubit system.
It is clear that when $N_{A}=1$ and $N_{B}>1$ or $N_{B}=1$ and $N_{A}>1$, then the concurrence decays to zero in the long time limit.
On the other hand,
as $N_{A}$, $N_{B}\rightarrow \infty$, the concurrence approaches to its initial value, i.e. 2$|C_{j}^{A}(0) C_{l}^{B}(0)|$. Also, if we let, for instance, $N_{B}\rightarrow \infty$ and rename $N_{A}\equiv N$ then the Eq. (24) becomes
\begin{eqnarray}
  C(\rho_{j,l})=2 \frac{N-1}{N} |C_{j}^{A}(0) C_{l}^{B}(0)|.
\end{eqnarray}
The concurrence evolution in (23) can be evaluated for different non-Markovian behaviors of the related subsystems as shown in Fig. 5 with $N_{A}=N_{B}=6$. At the long time limit, it is interesting to note that, apart from different  non-Markovian nature of the subsystems, the asymptotic concurrences are completely coincided to each others.

\subsection{W-type entanglement}
In this section, we extend the results of the previous section to three subsystems $A$, $B$ and $C$ contained in three independent Lorenzian reservoirs (see Fig. 1(c)).
Each subsystem has $N_{A}$, $N_{B}$ and $N_{C}$ qubits respectively. A W-type entangled state of $j^{th}$ qubit of subsystem $A$, $l^{th}$ qubit of subsystem $B$ and $m^{th}$ qubit of subsystem $C$, at time $t=0$, can be written as follows
\begin{eqnarray}
  |W\rangle_{ABC}=C_{j}^{A}(0)|e,g,g\rangle+C_{l}^{B}(0)|g,e,g\rangle+C_{m}^{C}(0)|g,g,e\rangle.
\end{eqnarray}
As shown in appendix B, in the standard computational basis as $\{|1\rangle \equiv|e,e,e\rangle, |2\rangle \\ \equiv|e,e,g\rangle, |3\rangle \equiv|e,g,e\rangle,|4\rangle \equiv|e,g,g\rangle, |5\rangle \equiv|g,e,e\rangle, |6\rangle \equiv|g,e,g\rangle, |7\rangle \equiv|g,g,e\rangle, |8\rangle \equiv|g,g,g\rangle\}$, the matrix elements of the three-qubit density operator at time $t$ are

\begin{eqnarray}
\begin{array}{c}
\rho_{44}(t)=|G^{A}(t)|^{2} |C_{j}^{A}(0)|^2,\\\\
\rho_{66}(t)=|G^{B}(t)|^{2} |C_{l}^{B}(0)|^2,\\\\
\rho_{77}(t)=|G^{C}(t)|^{2} |C_{m}^{C}(0)|^2,\\\\
\rho_{88}(t)=1-|G^{A}(t)|^{2}|C_{j}^{A}(0)|^2-|G^{B}(t)|^{2}|C_{l}^{B}(0)|^2-|G^{C}(t)|^{2}|C_{m}^{C}(0)|^2,\\\\
\rho_{46}(t)=\rho_{64}^{*}(t)=G^{A}(t) {G^{B}}^{*}(t) C_{j}^{A}(0) {C_{l}^{B}}^{*}(0),\\\\
\rho_{47}(t)=\rho_{74}^{*}(t)=G^{A}(t) {G^{C}}^{*}(t) C_{j}^{A}(0) {C_{m}^{C}}^{*}(0),\\\\
\rho_{67}(t)=\rho_{76}^{*}(t)=G^{B}(t) {G^{C}}^{*}(t) C_{l}^{B}(0) {C_{m}^{C}}^{*}(0),
\end{array}
\end{eqnarray}
and zero for the remained ones and $G^{A}(t), G^{B}(t)$ and $G^{C}(t)$ are similar to the case of Eq. (16). The parameter $\lambda_{A}$ ($\lambda_{B}$ and $\lambda_{C}$) is the spectral width of the coupling of $N_{A}$ ($N_{B}$ and $N_{C}$) qubits to its respective reservoir. Also, we assume that the coupling constants of the subsystems to their respective reservoirs are equal $\gamma_{A}=\gamma_{B}=\gamma_{C}=\gamma_{0}$.

To assess to the degree of tripartite entanglement analytically, the so-called lower bound of concurrence ($LBC$) is used in this way. Any
separable states have a vanishing $LBC$ but its inverse is not true. However, a positive
$LBC$ indicates the detection of entanglement with certainty yet. Thus, using of $LBC$ for evaluating the entanglement dynamics is acceptable. According to Ref. $\cite{Fei}$, the $LBC$ for a three-qubit state $\rho$ is defined as
\begin{eqnarray}
  LBC(\rho)=\sqrt{\frac{1}{3}\sum_{r=1}^{6} \Big\{\big[C_{r}^{(12|3)}(\rho)\big]^2+\big[C_{r}^{(23|1)}(\rho)\big]^2+\big[C_{r}^{(31|2)}(\rho)\big]^2\Big\}},
\end{eqnarray}
where
\begin{eqnarray}
  C_{r}^{(uv|w)}(\rho)=\mathrm{max}\Big\{0,\sqrt{\lambda_{r, s}^{(uv|w)}}-\sum_{t>s} \sqrt{\lambda_{r, t}^{(uv|w)}}\Big\}.
\end{eqnarray}
In Eq. (29), $\lambda_{r, t}^{(uv|w)}$ are the eigenvalues of the density matrix $\rho(L_{r}^{uv}\otimes \sigma_{y}^{w})\rho^*(L_{r}^{uv}\otimes \sigma_{y}^{w})$ in decreasing order where $L_{r}^{uv}(r=1, 2, ..., 6)$ are the six generators of the $SO(4)$ group acting on the qubits $u$ and $v$, and $\sigma_{y}^{w}$ is the y-component Pauli matrix acting on the qubit $w$.
For the reduced three-qubit density matrix of our concern, $LBC$ can be obtained as
\begin{eqnarray}
\begin{array}{c}
   LBC(\rho_{j, l, m})=\sqrt{\frac{8}{3}}\big(|G^{A}(t) G^{B}(t)|^2|C_{j}^{A}(0) C_{l}^{B}(0)|^2+|G^{A}(t) G^{C}(t)|^2|C_{j}^{A}(0) C_{m}^{C}(0)|^2+ \\\\
  |G^{B}(t) G^{C}(t)|^2|C_{l}^{B}(0) C_{m}^{C}(0)|^2\big)^{\frac{1}{2}}.
\end{array}
 \end{eqnarray}
Fig. 6, shows the $LBC$ for the three-qubit in Markovian and non-Markovian regimes.
Notice that similar to the previous section, for $N_{A}=N_{B}=N_{C}=1$ (in the absence of additional qubits in the subsystems A, B and C), the $LBC$ eventually
decays to zero but there is a steady value for $LBC$ retained by increasing $N_{A}$, $N_{B}$ and $N_{C}$ in both Markovian and non-Markovian regimes.
In the limit $t\rightarrow \infty$, the $LBC$ reduces to
\begin{eqnarray}
  \begin{array}{c}
    LBC(\rho_{j, l, m})=\sqrt{\frac{8}{3}}\big(\frac{(N_{A}-1)^{2}(N_{B}-1)^{2}}{N_{A}^{2}N_{B}^{2}}|C_{j}^{A}(0) C_{l}^{B}(0)|^2+
  \frac{(N_{A}-1)^{2}(N_{C}-1)^{2}}{N_{A}^{2}N_{C}^{2}}|C_{j}^{A}(0) C_{m}^{C}(0)|^2+\\\\
    \frac{(N_{B}-1)^{2}(N_{C}-1)^{2}}{N_{B}^{2}N_{C}^{2}}|C_{l}^{B}(0) C_{m}^{C}(0)|^2\big)^{\frac{1}{2}}.
  \end{array}
\end{eqnarray}
Consider, for example, $N_{C}=1$ then the $LBC$ in (31) becomes
\begin{eqnarray}
      LBC(\rho_{j, l, m})=\sqrt{\frac{8}{3}}\frac{(N_{A}-1)(N_{B}-1)}{N_{A}N_{B}}|C_{j}^{A}(0) C_{l}^{B}(0)|,
  \end{eqnarray}
and in the same way, if $N_{B}=1$ and $N_{C}=1$, then $LBC$ is equal to zero.
Ultimately, in the limits $N_{A},N_{B},N_{C}\rightarrow \infty$, the $LBC$ reaches
\begin{eqnarray}
  LBC(\rho_{j, l, m})=\sqrt{\frac{8}{3}}\big(|C_{j}^{A}(0) C_{l}^{B}(0)|^2+|C_{j}^{A}(0) C_{m}^{C}(0)|^2+|C_{l}^{B}(0) C_{m}^{C}(0)|^2\big)^{\frac{1}{2}},
\end{eqnarray}
and also for the $LBC$ in (32), as $N_{A}, N_{B}\rightarrow \infty$, we have
\begin{eqnarray}
      LBC(\rho_{j, l, m})=\sqrt{\frac{8}{3}}|C_{j}^{A}(0) C_{l}^{B}(0)|.
\end{eqnarray}

The $LBC$ evolution in (30) can be evaluated under different non-Markovian behaviors of the related subsystems as shown
in Fig. 7 with $N_{A}=N_{B}=N_{C}=6$. Obviously, at long time limit, it is interesting to note that, apart from
different non-Markovian behaviors of the subsystems, the asymptotic $LBC$s are completely coincided to each others again.

\section{Conclusion}
We investigated the preservation of quantum coherence of a single-qubit interacting with a zero-temperature thermal reservoir through the addition of non-interacting qubits into the reservoir.
Also, we discussed the extension of this scheme for entanglement protection of two
and three distant non-interacting qubits, each of which individually has been contained in an independent reservoir.
At limit $t\rightarrow\infty$, explicit dependence of the coherence measure, bipartite and tripartite concurrences on the number of additional qubits were derived.
It was pointed out that, by increasing the number of additional qubits in each
reservoir, the initial coherence and the respective entanglements are completely protected in both
Markovian and non-Markovian regimes.
Interestingly it was shown that, for preserving of initial state entanglement, the dynamics of each subsystem is not necessarily similar to dynamics of the other subsystem (subsystems) from Markovian and non-Markovian point of views. On the other hand, from the experimental point of view, the proposed scheme can be realized using lossy (imperfect) cavities. As illustrated, for preserving the entanglement, it is not important for the cavities to have the same spectral density. It should be noted that the scheme can be extended for protecting higher order multipartite entanglement of qubits distantly contained in the Lorentzian reservoirs such as cavities. Finally, the proposed scheme of this paper can be extended for investigating the possibility of protection of coherence and entanglement against the temperature of the reservoir through the additional qubits which can be regarded as the subject of future research.

\newpage
\vspace{1cm} \setcounter{section}{0}
 \setcounter{equation}{0}
 \renewcommand{\theequation}{A.\arabic{equation}}
{\Large{Appendix A:}}\\
\textbf{Details of derivation of Eq. (13)}:
Taking the Laplace transform from both sides of Eq. (10) gives the following set of equations
\begin{eqnarray}
  p C_{j}(p)-C_{j}(0)=-\beta_{j} \mathcal{L}\{f(t)\} \sum_{l=1}^{N} \beta_{l} C_{l}(p),
\end{eqnarray}
where $j=1,2,...,N$. Here we use the notation $F(p)=\mathcal{L} \{F(t)\}=\int_{0}^{\infty} F(t) e^{-pt} dt$.
By dividing Eq. (A.1) to $\beta_{j}$, we observe that the right hand sides of $N$ equations are equal so the following relation between the coefficients is obtained
\begin{eqnarray}
  \frac{p C_{1}(p)-C_{1}(0)}{\beta_{1}}=\frac{p C_{2}(p)-C_{2}(0)}{\beta_{2}}=...=\frac{p C_{j}(p)-C_{j}(0)}{\beta_{j}}=...=\frac{p C_{N}(p)-C_{N}(0)}{\beta_{N}}.
\end{eqnarray}
By writing the coefficients $C_{l}(p)$ ($l\neq j$) in terms of $C_{j}(p)$ and inserting them into the Eq. (A.1), we have
\begin{eqnarray}
      p C_{j}(p)-C_{j}(0)=-\mathcal{L}\{f(t)\} \bigg{(}\beta_{j}^2 C_{j}(p)+\frac{1}{p}\sum_{l\neq j}^{N} \Big{[} \beta_{l}^2(pC_{j}(p)-C_{j}(0))+\beta_{j}\beta_{l}C_{l}(0) \Big{]} \bigg{)},
\end{eqnarray}
For a Lorentzian spectral density given by Eq. (12), the correlation function $f(t)$ can be calculated by using Eq. (11), therefore we have
\begin{eqnarray}
  f(t)=\frac{\gamma_{0} \lambda}{2}e^{-(\lambda-i\Delta)t},
\end{eqnarray}
and its Laplace transform is written as
\begin{eqnarray}
  \mathcal{L}\{f(t)\}=\frac{\gamma_{0} \lambda}{2(p+\lambda -i\Delta)},
\end{eqnarray}
After substituting Eq. (A.5) into Eq. (A.3), the coefficients $C_{j}(p)$ can be obtained as
\begin{eqnarray}
      C_{j}(p)=\frac{2(p+\lambda -i\Delta)}{2(p+\lambda -i\Delta)+ \sum_{l=1}^{N} \beta_{l}^2}C_{j}(0)+\frac{\gamma_{0} \lambda \big{(}\sum_{l\neq j}^{N} \big{[} \beta_{l}^2 C_{j}(0)-\beta_{j}\beta_{l} C_{j}(0)\big{]} \big{)}}{p \Big{(} 2(p+\lambda -i\Delta)+ \sum_{l=1}^{N} \beta_{l}^2\Big{)}},
\end{eqnarray}
Finally, the inverse Laplace transform of $C_{j}(p)$ gives Eq. (13).

\vspace{1cm} \setcounter{section}{0}
 \setcounter{equation}{0}
 \renewcommand{\theequation}{B.\arabic{equation}}
{\Large{Appendix B:}}\\
\textbf{Dynamics of two and three independent qubits}:
As discussed in Ref. $\cite{Bellomo}$, a single-qubit dynamics has the form
\begin{eqnarray}
\rho_{mm'}(t)=\sum_{nn'} \chi^{nn'}_{mm'}(t) \rho_{nn'}(0),
\end{eqnarray}
where $\rho_{mm'}(t)=\langle m| \rho(t) |m'\rangle$ with $m,n=e,g$.
The matrix whose entries are the values $\chi^{nn'}_{mm'}(t)$ is said to form a matrix representation of $\chi(t)$. By using Eq. (17),
the matrix representation of $\chi(t)$ in the standard computational basis $\{|e,e\rangle, |e,g\rangle, |g,e\rangle, |g,g\rangle\}$ can be obtained as
\begin{eqnarray}
\chi=\left(
\begin{array}{cccc}
  |G(t)|^2 & 0 & 0 & 0 \\
  0 & G(t) & 0 & 0 \\
  0 & 0 & G(t)^{*} & 0 \\
  1-|G(t)|^2 & 0 & 0 & 1
  \end{array}
   \right),
\end{eqnarray}
where $\chi^{nn'}_{mm'}(t)=\langle m,m'| \chi(t) |n,n'\rangle$.

We now consider a system consisting of two independent qubits, each locally interacting with its own reservoir.
The complete dynamics of the two-qubit system can be obtained by knowing the single-qubit dynamics which has been obtained in Eq. (17).
Under these conditions, given the dynamics of each qubit as $\rho^{A}_{m_{1}m'_{1}}(t)=\sum_{n_{1}n'_{1}} A^{n_{1}n'_{1}}_{m_{1}m'_{1}}(t) \rho_{n_{1}n'_{1}}(0)$ and
$\rho^{B}_{m_{2}m'_{2}}(t)=\sum_{n_{2}n'_{2}} B^{n_{2}n'_{2}}_{m_{2}m'_{2}}(t) \rho_{n_{2}n'_{2}}(0)$,
the dynamics of the two-qubit system is expressed by
\begin{eqnarray}
\begin{array}{c}
\rho^{AB}_{m_{1}m'_{1},m_{2}m'_{2}}(t)=\sum_{n_{1}n'_{1}} \sum_{n_{2}n'_{2}} A^{n_{1}n'_{1}}_{m_{1}m'_{1}}(t) B^{n_{2}n'_{2}}_{m_{2}m'_{2}}(t) \rho^{AB}_{n_{1}n'_{1},n_{2}n'_{2}}(0)=\\\\
\sum_{n_{1}n'_{1}} \sum_{n_{2}n'_{2}} E^{n_{1}n'_{1},n_{2}n'_{2}}_{m_{1}m'_{1},m_{2}m'_{2}}(t) \rho^{AB}_{n_{1}n'_{1},n_{2}n'_{2}}(0),
\end{array}
\end{eqnarray}
where $\rho^{AB}_{m_{1}m'_{1},m_{2}m'_{2}}(t)=\langle m_{1},m_{2}| \rho^{AB}(t) |m'_{1},m'_{2}\rangle$ and $E^{n_{1}n'_{1},n_{2}n'_{2}}_{m_{1}m'_{1},m_{2}m'_{2}}(t)=
\langle m_{1}m'_{1},m_{2}m'_{2}| A\otimes B |n_{1}n'_{1},n_{2}n'_{2}\rangle$ with $m_{1},n_{1},m_{2},n_{2}=e,g$. Since the qubits are in general in different environments so that their evolution is characterized by the different functions $A(t)$ and $B(t)$ as
\begin{eqnarray}
\begin{array}{c}
A(t)=\left(
\begin{array}{cccc}
  |G^{A}(t)|^2 & 0 & 0 & 0 \\
  0 & G^{A}(t) & 0 & 0 \\
  0 & 0 & {G^{A}}^{*}(t) & 0 \\
  1-|G^{A}(t)|^2 & 0 & 0 & 1
  \end{array}
   \right),\\\\
   B(t)=\left(
             \begin{array}{cccc}
               |G^{B}(t)|^2 & 0 & 0 & 0 \\
               0 & G^{B}(t) & 0 & 0 \\
               0 & 0 & {G^{B}}^{*}(t) & 0 \\
               1-|G^{B}(t)|^2 & 0 & 0 & 1
   \end{array}
   \right),
\end{array}
\end{eqnarray}
where $G^{A}(t)$ and $G^{B}(t)$ can be considered as
\begin{eqnarray}
  G^{j}(t)=\frac{N_{j}-1}{N_{j}}+\frac{e^{-\Lambda_{j} t/2}}{N_{j}} \big(cosh{(\frac{D_{j}t}{2})}+\frac{\Lambda_{j}}{D_{j}} sinh{(\frac{D_{j}t}{2})}\big),
\end{eqnarray}
and $D_{j}=\sqrt{\Lambda_{j}^{2}-2\gamma_{j} \lambda_{j} N_{j}}$ with $j=A, B$.

In the following, we extend this procedure to explore the dynamics of three independent qubits, each locally interacting with its own reservoir.
By considering the dynamics of the third qubit as $\rho^{C}_{m_{3}m'_{3}}(t)=\sum_{n_{3}n'_{3}} C^{n_{3}n'_{3}}_{m_{3}m'_{3}}(t) \rho_{n_{1}n'_{1}}(0)$, the dynamics of the three-qubit system is simply given by
\begin{eqnarray}
\begin{array}{c}
\rho^{ABC}_{m_{1}m'_{1},m_{2}m'_{2},m_{3}m'_{3}}(t)=\sum_{n_{1}n'_{1}} \sum_{n_{2}n'_{2}} \sum_{n_{3}n'_{3}} A^{n_{1}n'_{1}}_{m_{1}m'_{1}}(t) B^{n_{2}n'_{2}}_{m_{2}m'_{2}}(t) C^{n_{3}n'_{3}}_{m_{3}m'_{3}}(t) \rho^{ABC}_{n_{1}n'_{1},n_{2}n'_{2},n_{3}n'_{3}}(0)=\\\\
\sum_{n_{1}n'_{1}} \sum_{n_{2}n'_{2}} \sum_{n_{3}n'_{3}} F^{n_{1}n'_{1},n_{2}n'_{2},n_{3}n'_{3}}_{m_{1}m'_{1},m_{2}m'_{2},m_{3}m'_{3}}(t) \rho^{ABC}_{n_{1}n'_{1},n_{2}n'_{2},n_{3}n'_{3}}(0),
\end{array}
\end{eqnarray}
where $\rho^{ABC}_{m_{1}m'_{1},m_{2}m'_{2},m_{3}m'_{3}}(t)=\langle m_{1},m_{2},m_{3}| \rho^{ABC}(t) |m'_{1},m'_{2},m'_{3}\rangle$ and also $F^{n_{1}n'_{1},n_{2}n'_{2},n_{3}n'_{3}}_{m_{1}m'_{1},m_{2}m'_{2},m_{3}m'_{3}}(t)=
\langle m_{1}m'_{1},m_{2}m'_{2},m_{3}m'_{3}| A \otimes B \otimes C |n_{1}n'_{1},n_{2}n'_{2},n_{3}n'_{3}\rangle$ with $m_{1},n_{1},m_{2},n_{2},m_{3},n_{3}=e,g$.
By knowing the matrix representations of $A(t)$ and $B(t)$ in Eq. (B.4), the function $C(t)$ is obtained as
\begin{eqnarray}
\begin{array}{c}
C(t)=\left(
\begin{array}{cccc}
  |G^{C}(t)|^2 & 0 & 0 & 0 \\
  0 & G^{C}(t) & 0 & 0 \\
  0 & 0 & {G^{C}}^{*}(t) & 0 \\
  1-|G^{C}(t)|^2 & 0 & 0 & 1
  \end{array}
   \right),
\end{array}
\end{eqnarray}
where $G^{A}(t)$, $G^{B}(t)$ and $G^{C}(t)$ are as
\begin{eqnarray}
  G^{j}(t)=\frac{N_{j}-1}{N_{j}}+\frac{e^{-\Lambda_{j} t/2}}{N_{j}} \big(cosh{(\frac{D_{j}t}{2})}+\frac{\Lambda_{j}}{D_{j}} sinh{(\frac{D_{j}t}{2})}\big),
\end{eqnarray}
and $D_{j}=\sqrt{\Lambda_{j}^{2}-2\gamma_{j} \lambda_{j} N_{j}}$ with $j=A, B, C$.

\newpage
Fig. 1. (a) The system of $N$ noninteracting qubits contained in a common reservoir. (b) Composite system consisting of two subsystems $A$ and $B$ each of which contained in an independent reservoir with $N_{A}$ and $N_{B}$ qubits respectively. (c) Composite system consisting of three subsystems $A$, $B$ and $C$ each of which contained in an independent reservoir with $N_{A}$, $N_{B}$ and $N_{C}$ qubits respectively.

\begin{figure}
\centering
    \centering
        {
        \includegraphics[width=2.6in]{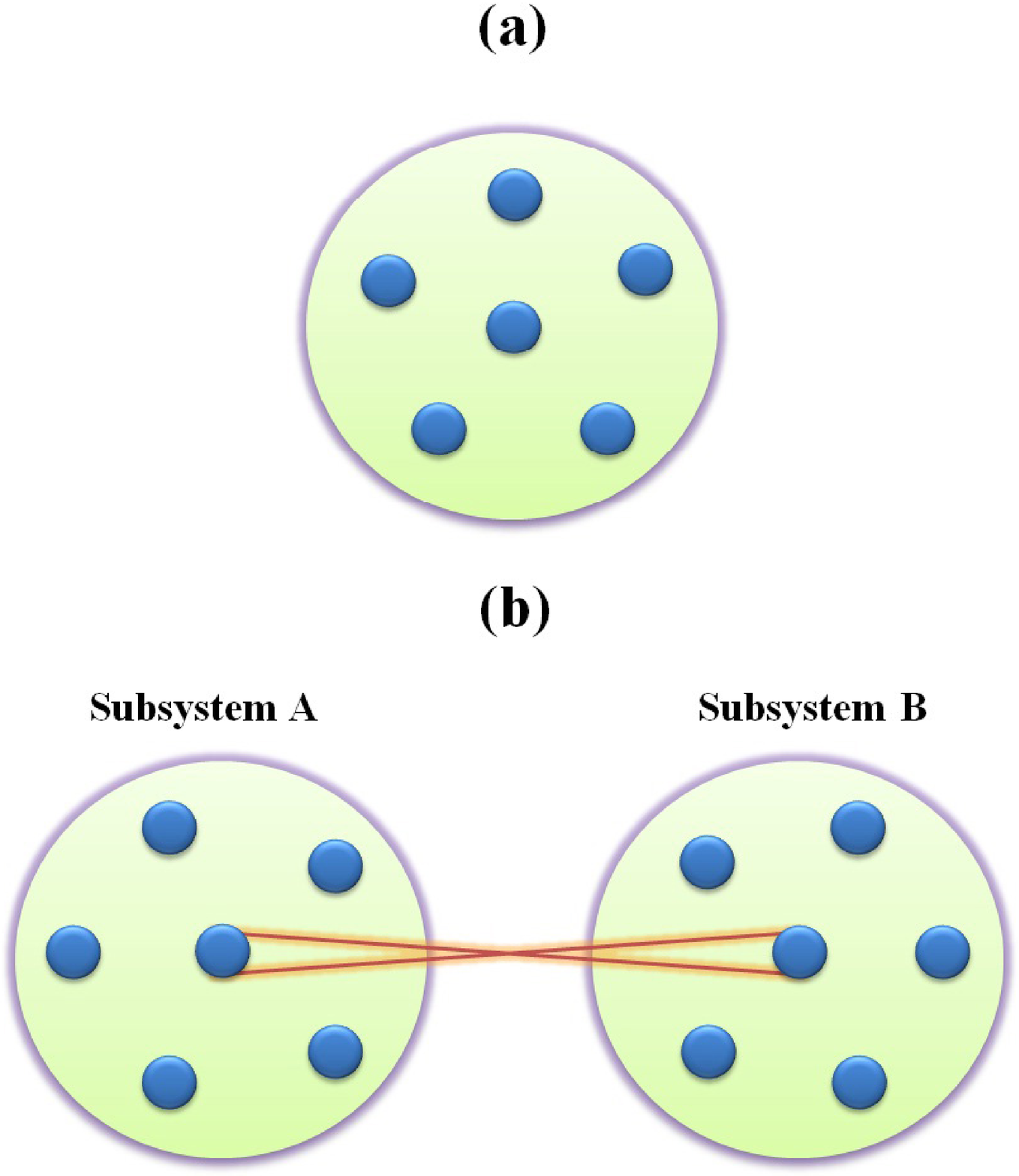}
        \label{fig:first_sub}
    }
    \\
    \centering
       {
        \includegraphics[width=2.6in]{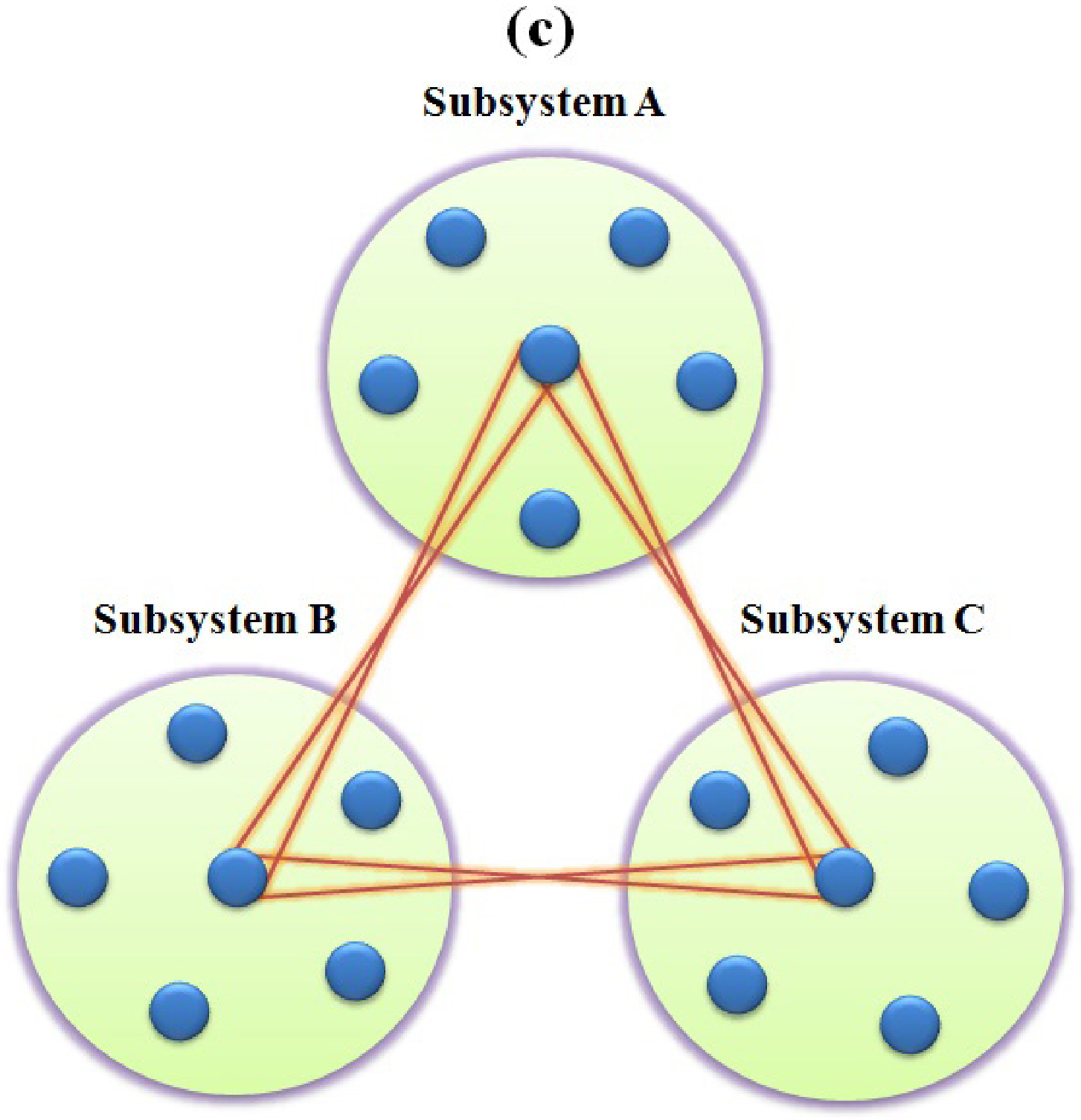}
        \label{fig:second_sub}
    }\\
    \centering
    \caption{}
\end{figure}

\newpage
Fig. 2. The Coherence measure $\xi(\rho_{j}(t))$ in terms of $\gamma_{0} t$ with $N=1$ (solid line), $N=2$ (dashed line), $N=3$ (dotted line) and $N=6$ (dotted dashed line). (a) and (b) show the coherence behavior in Markovian regime with $\lambda=15 \gamma_{0}$ and, (c) and (d) in non-Markovian regime with $\lambda=0.5 \gamma_{0}$. The $j^{th}$ qubit has been initially prepared in the state $\frac{1}{\sqrt{2}}|g\rangle+\frac{1}{\sqrt{2}}|e\rangle$. Panels (a) and (c) display the coherence without detuning ($\Delta=0$) and (b) and (d) display with detuning ($\Delta=2$).

\begin{figure}
        \qquad \qquad\qquad\qquad \qquad a \qquad\qquad \quad\qquad\qquad\qquad\qquad\qquad\qquad b\\{
        \includegraphics[width=3in]{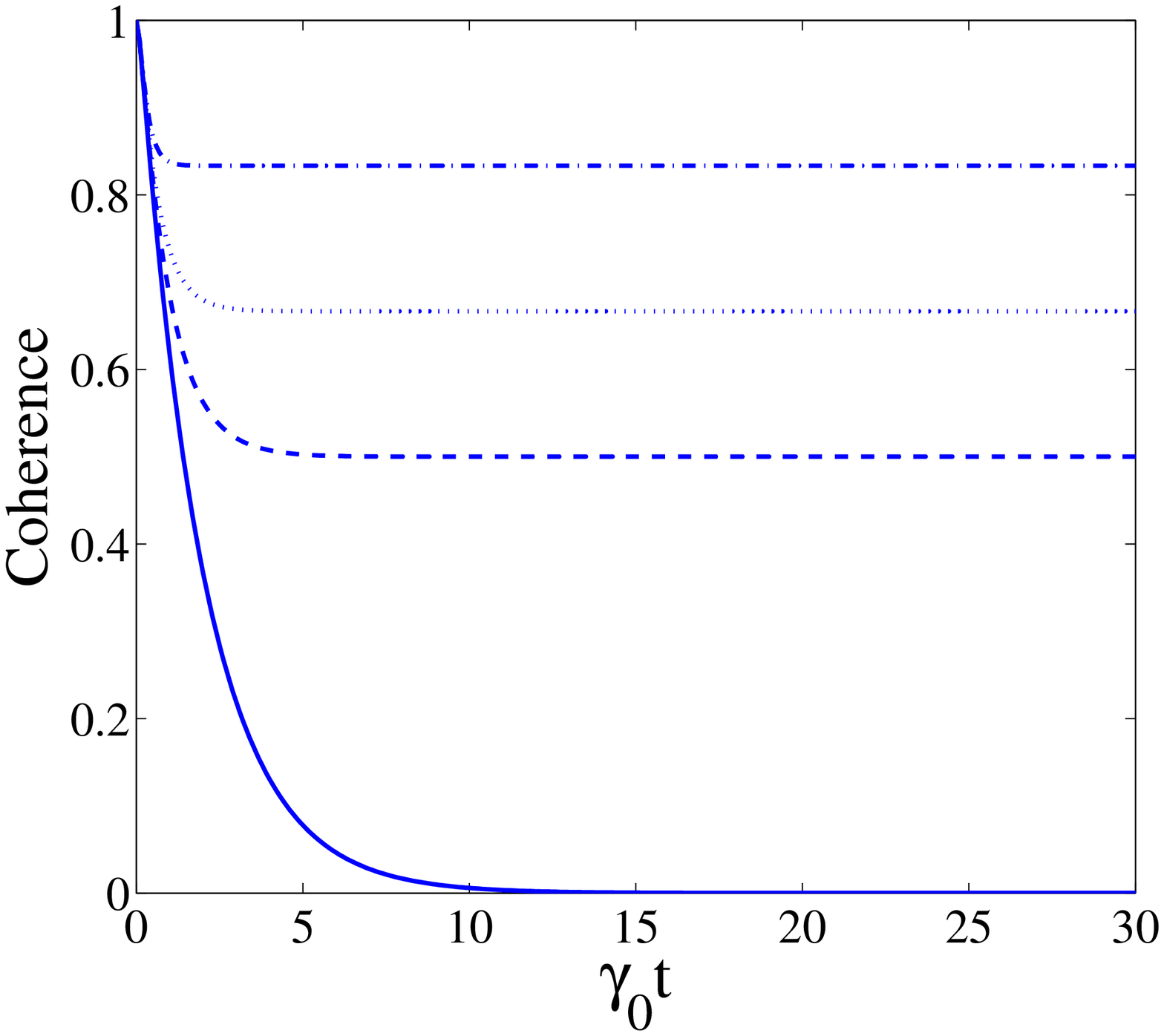}
        \label{fig:first_sub}
    }{
        \includegraphics[width=3in]{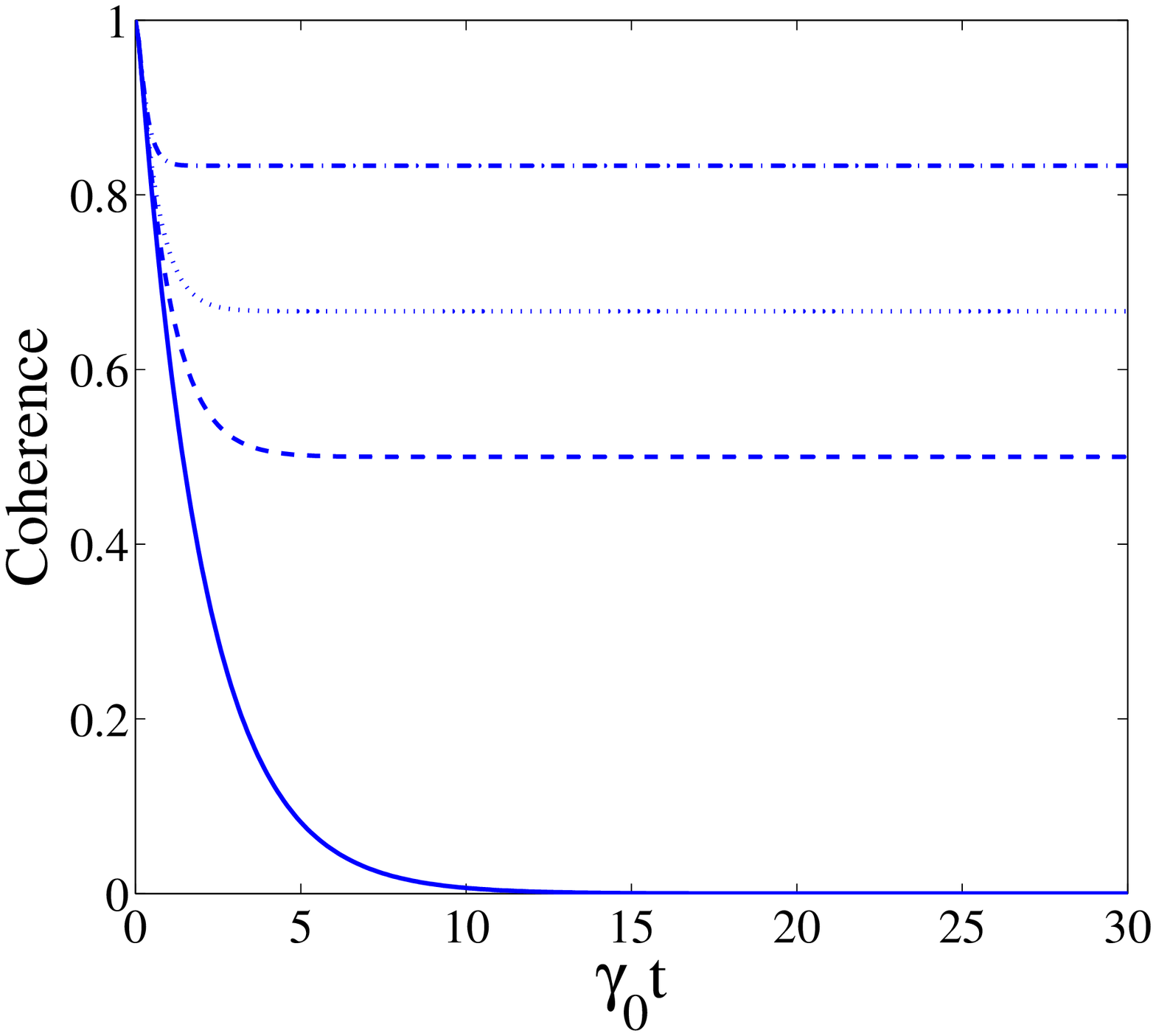}
        \label{fig:second_sub}
    }\\ \par \quad \quad\qquad\qquad\qquad \qquad c \qquad \qquad\qquad\qquad\qquad\quad\qquad\qquad\qquad d\\{
        \includegraphics[width=3in]{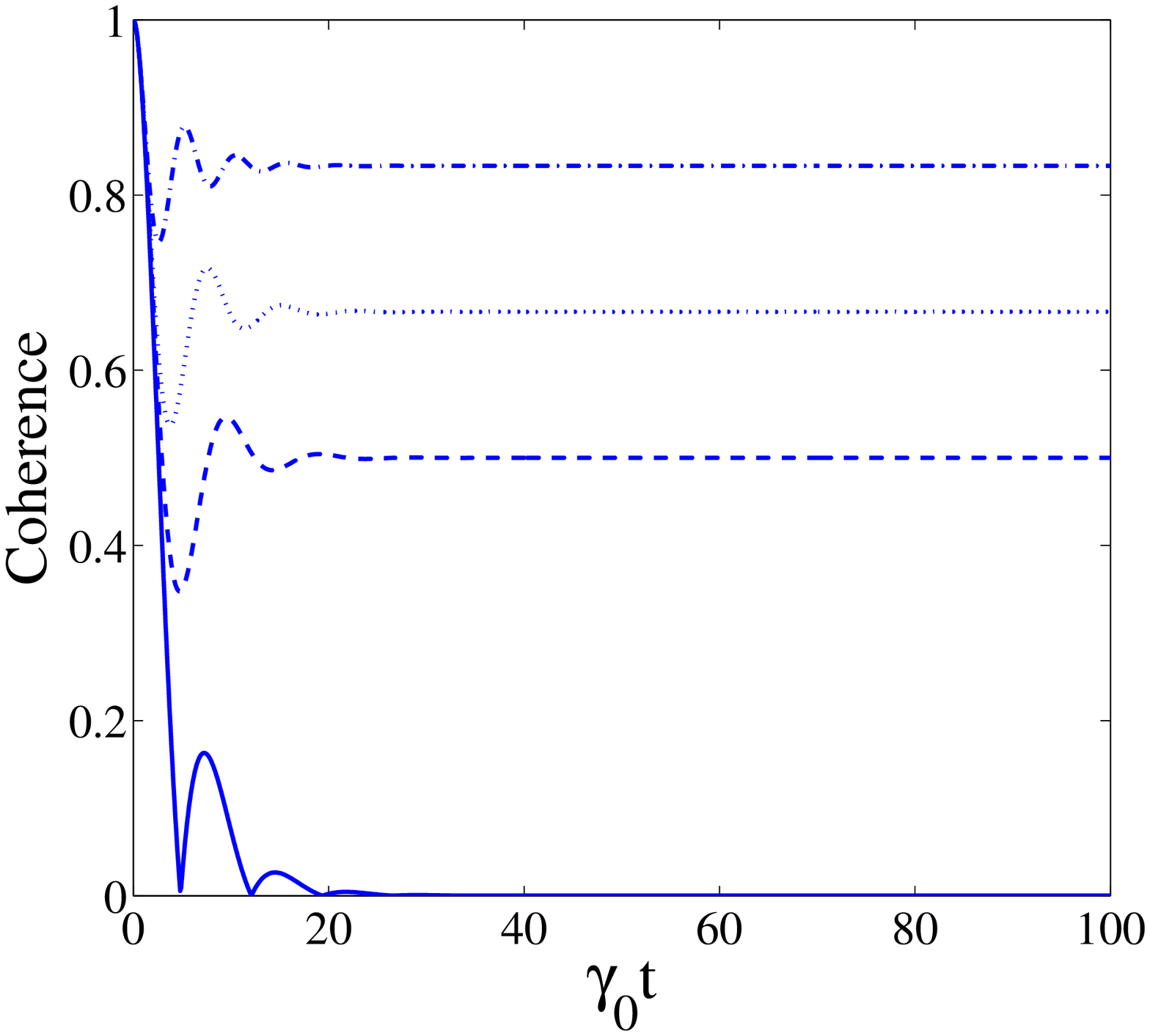}
        \label{fig:first_sub}
    }{
        \includegraphics[width=3in]{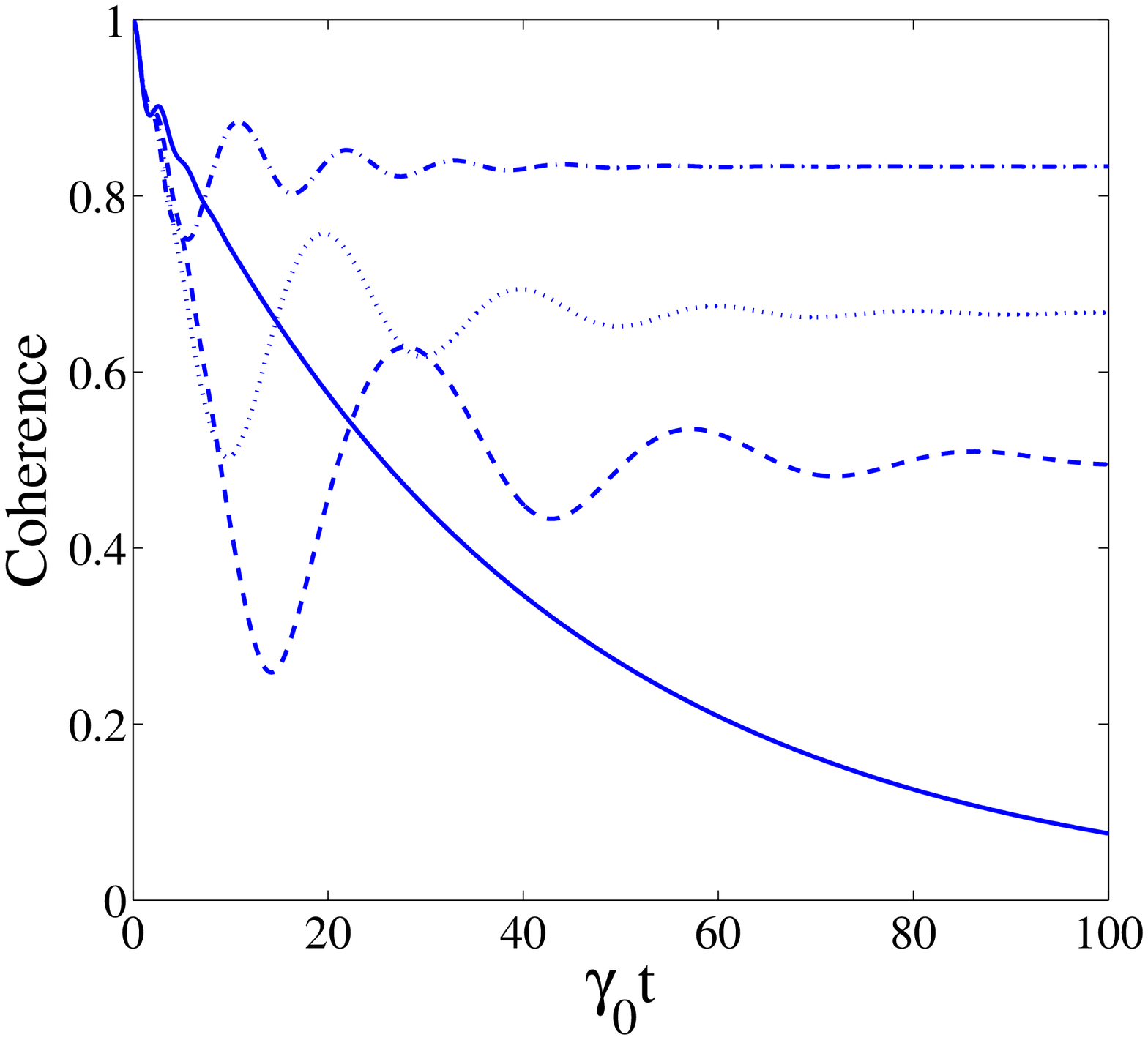}
        \label{fig:second_sub}
    }
    \caption{}
    \end{figure}

\newpage
Fig. 3. The behaviors of decay rate $\Gamma(t)$ in terms of $\gamma_{0} t$ with $N=1$ (solid line), $N=2$ (dashed line), $N=3$ (dotted line) and $N=6$ (dotted dashed line) where (a) and (b) for Markovian regime with $\lambda=15 \gamma_{0}$ and, (c) and (d) for non-Markovian regime with $\lambda=0.5 \gamma_{0}$. The $j^{th}$ qubit has been initially prepared in the state $\frac{1}{\sqrt{2}}|g\rangle+\frac{1}{\sqrt{2}}|e\rangle$. Panels (a) and (c) display the coherence without detuning ($\Delta=0$) and, (b) and (d) display with detuning ($\Delta=2$).

\begin{figure}
        \qquad \qquad\qquad\qquad \qquad a \qquad\qquad \quad\qquad\qquad\qquad\qquad\qquad\qquad b\\{
        \includegraphics[width=3in]{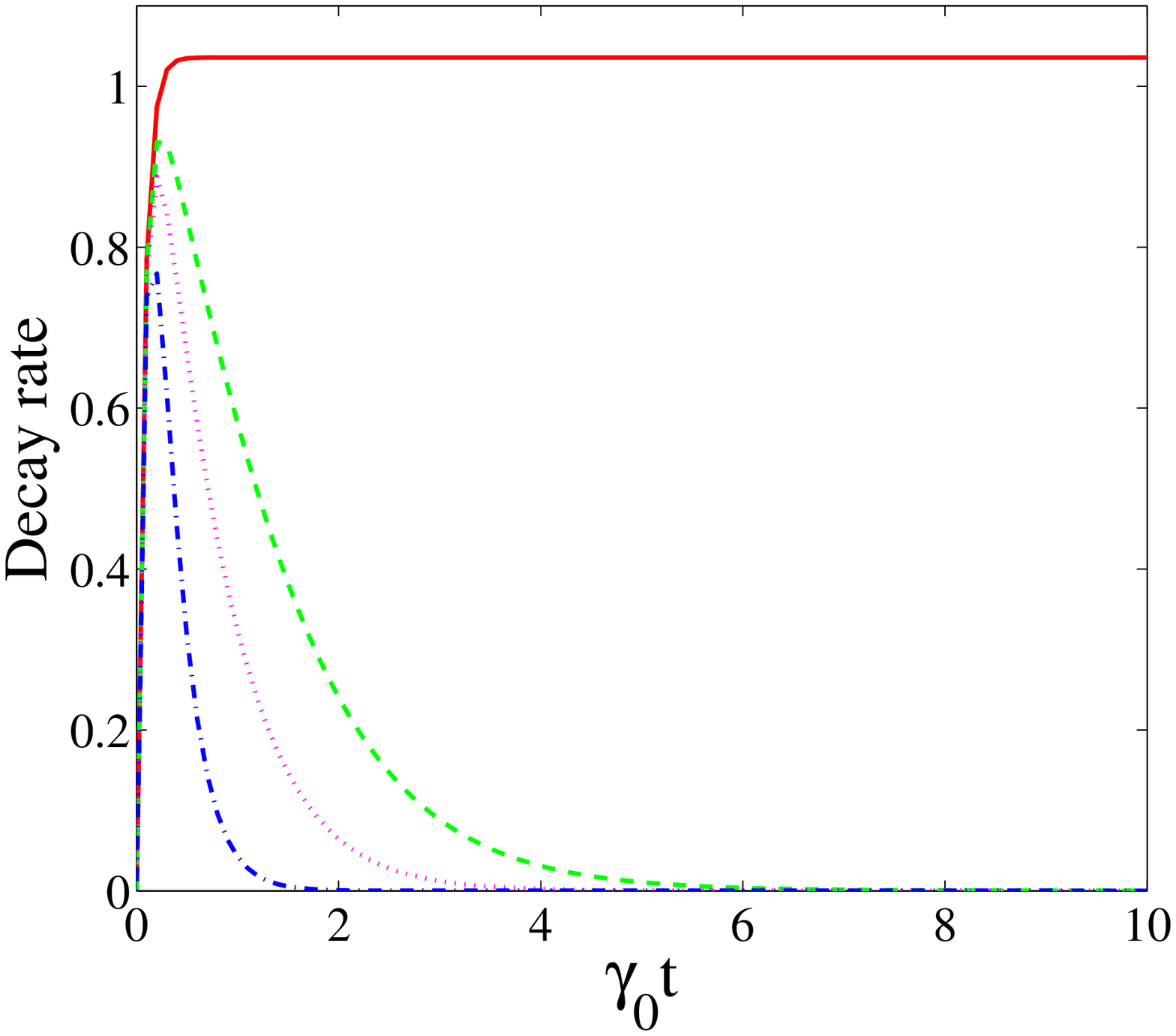}
        \label{fig:first_sub}
    }{
        \includegraphics[width=3in]{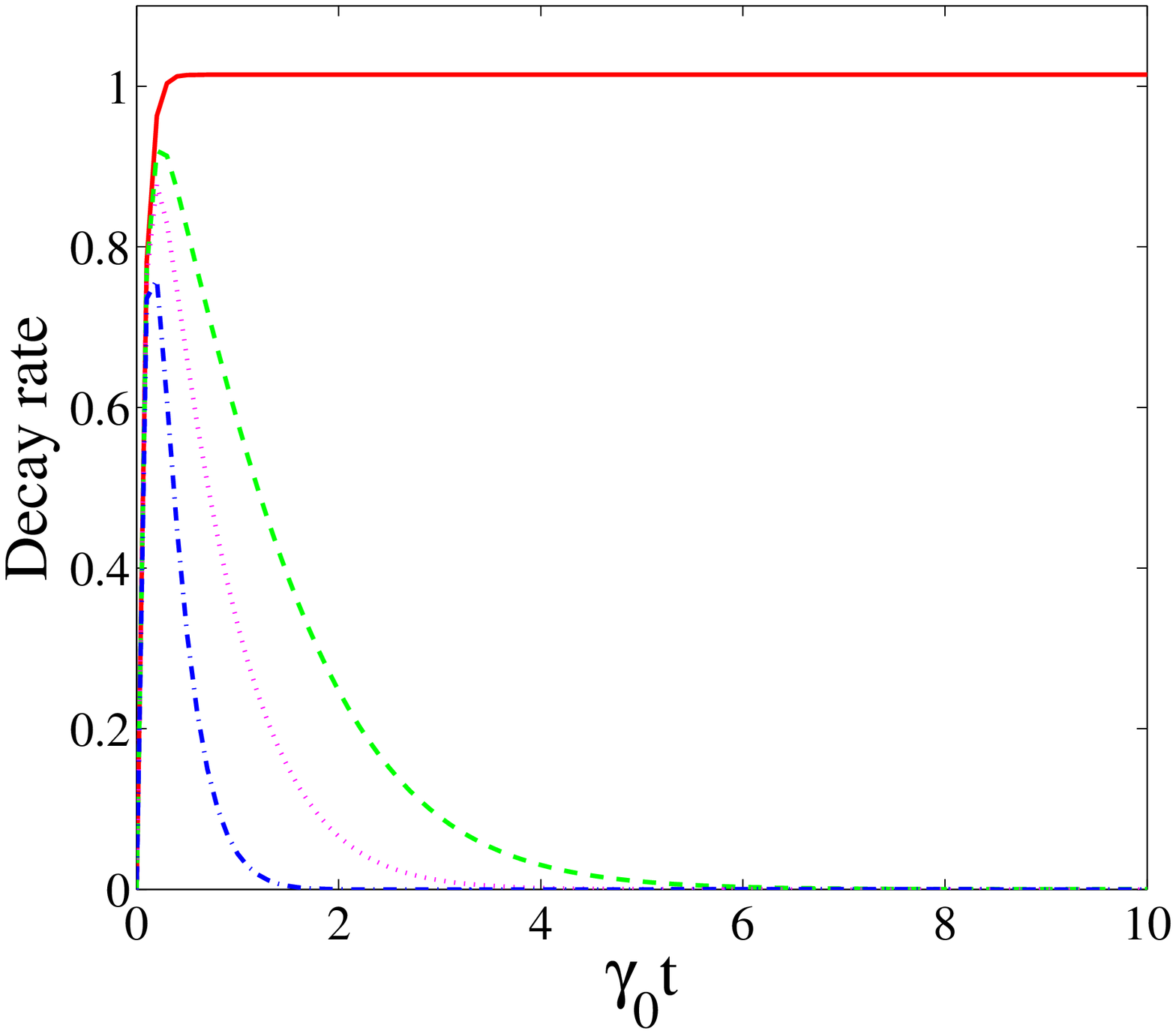}
        \label{fig:second_sub}
    }\\ \par \quad \quad\qquad\qquad\qquad \qquad c \qquad \qquad\qquad\qquad\qquad\quad\qquad\qquad\qquad d\\{
        \includegraphics[width=3in]{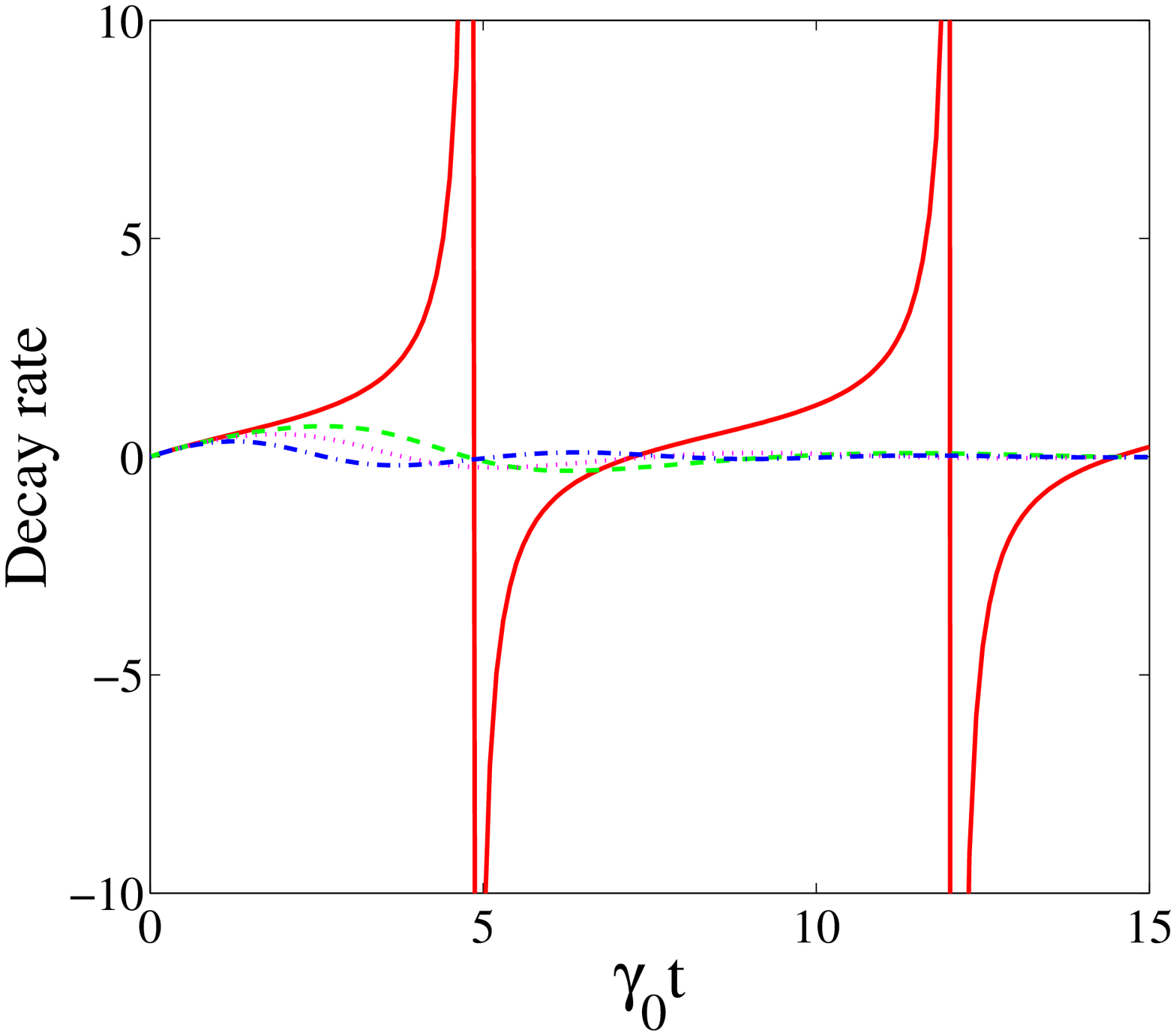}
        \label{fig:first_sub}
    }{
        \includegraphics[width=3in]{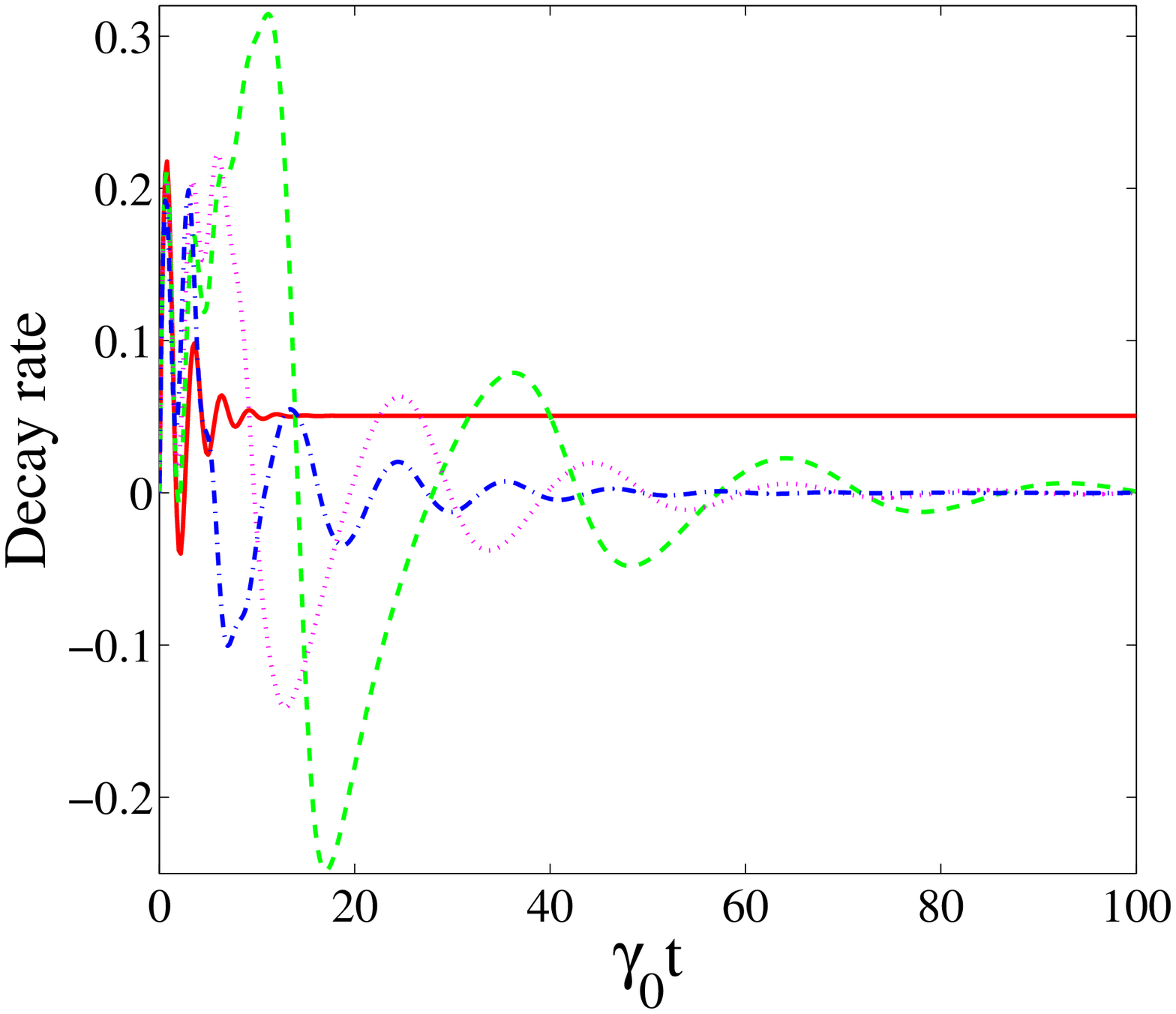}
        \label{fig:second_sub}
    }
    \caption{}
    \end{figure}

\newpage
Fig. 4. Concurrence as a function of $\gamma_{0} t$ with $\Delta_{A}=\Delta_{B}=2$, (a) Markovian regime with $\lambda_{A}=\lambda_{B}=15 \gamma_{0}$
and (b) non-Markovian regime with $\lambda_{A}=\lambda_{B}=0.5 \gamma_{0}$.
The initial state entanglement is determined by $C_{j}^{A}(0)=C_{l}^{B}(0)=\frac{1}{\sqrt{2}}$ with $C_{i}^{A}(0)=0$ for  $i\neq j$ and $C_{k}^{B}(0)=0$ for $k\neq l$.

\begin{figure}
        \qquad \qquad\qquad\qquad \qquad a \qquad\qquad \qquad\quad\qquad\qquad\qquad\qquad\quad\qquad\qquad\qquad b\\{
        \includegraphics[width=3.4in]{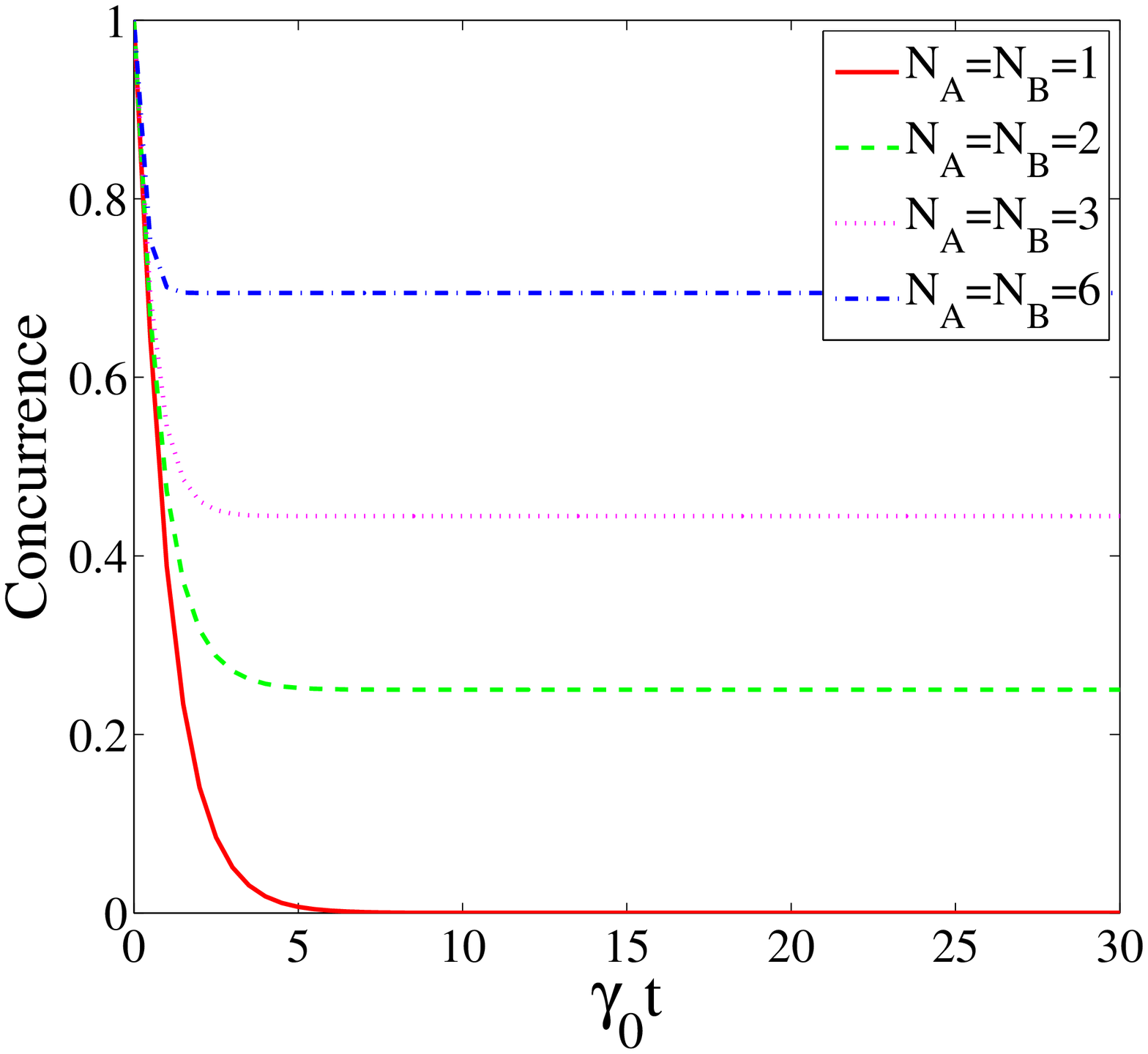}
        \label{fig:first_sub}
    }{
        \includegraphics[width=3.4in]{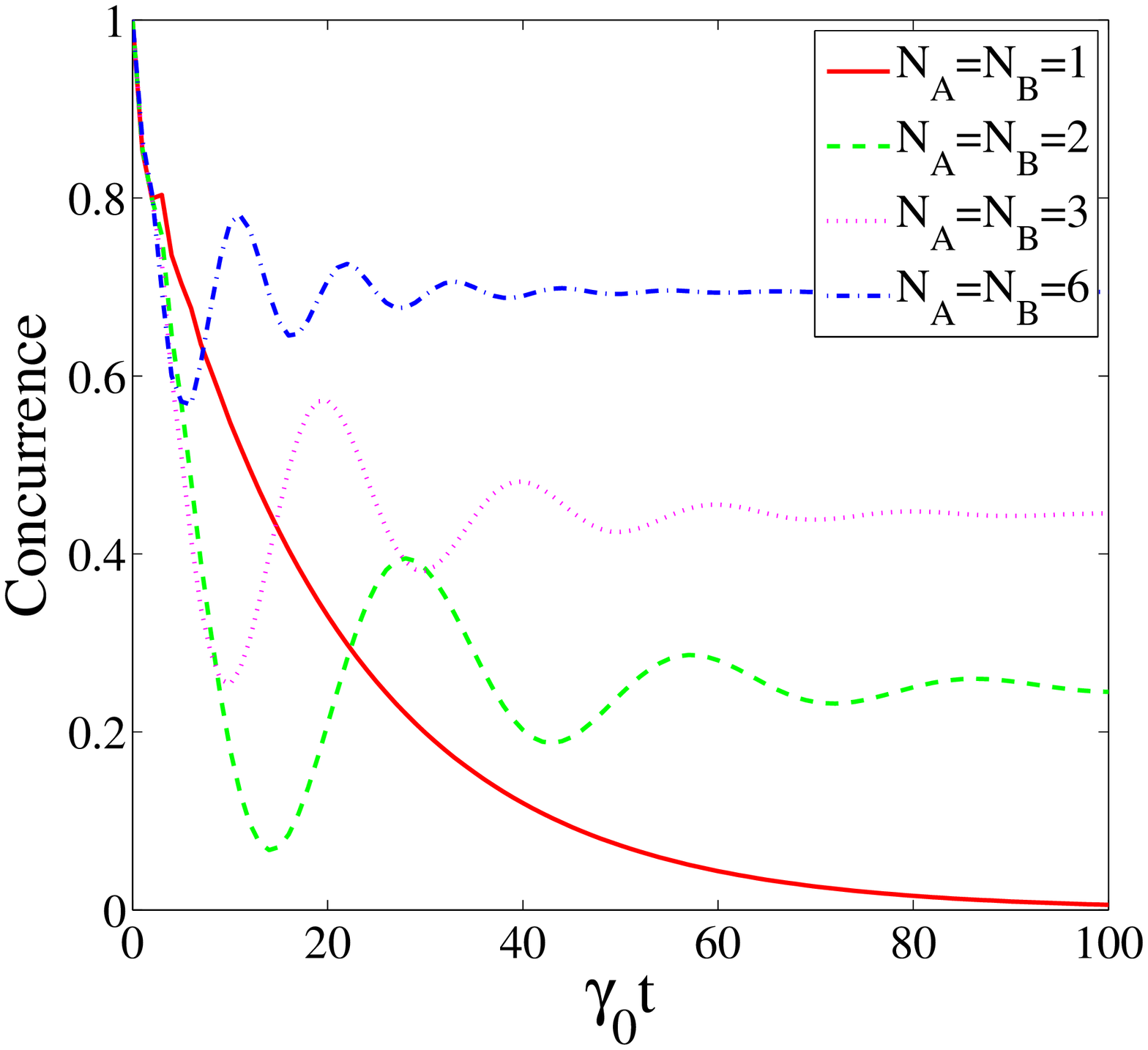}
        \label{fig:second_sub}
    }
    \caption{}
    \end{figure}

\newpage
Fig. 5. Concurrence as a function of $\gamma_{0} t$ with $N_{A}=N_{B}=6$ and $\Delta_{A}=\Delta_{B}=2$. For Markovian regime (M), $\lambda_{j}=15 \gamma_{0}$ and for non-Markovian regime (N), $\lambda_{j}=0.5 \gamma_{0}$.

\begin{figure}
\centering
\includegraphics[width=300 pt]{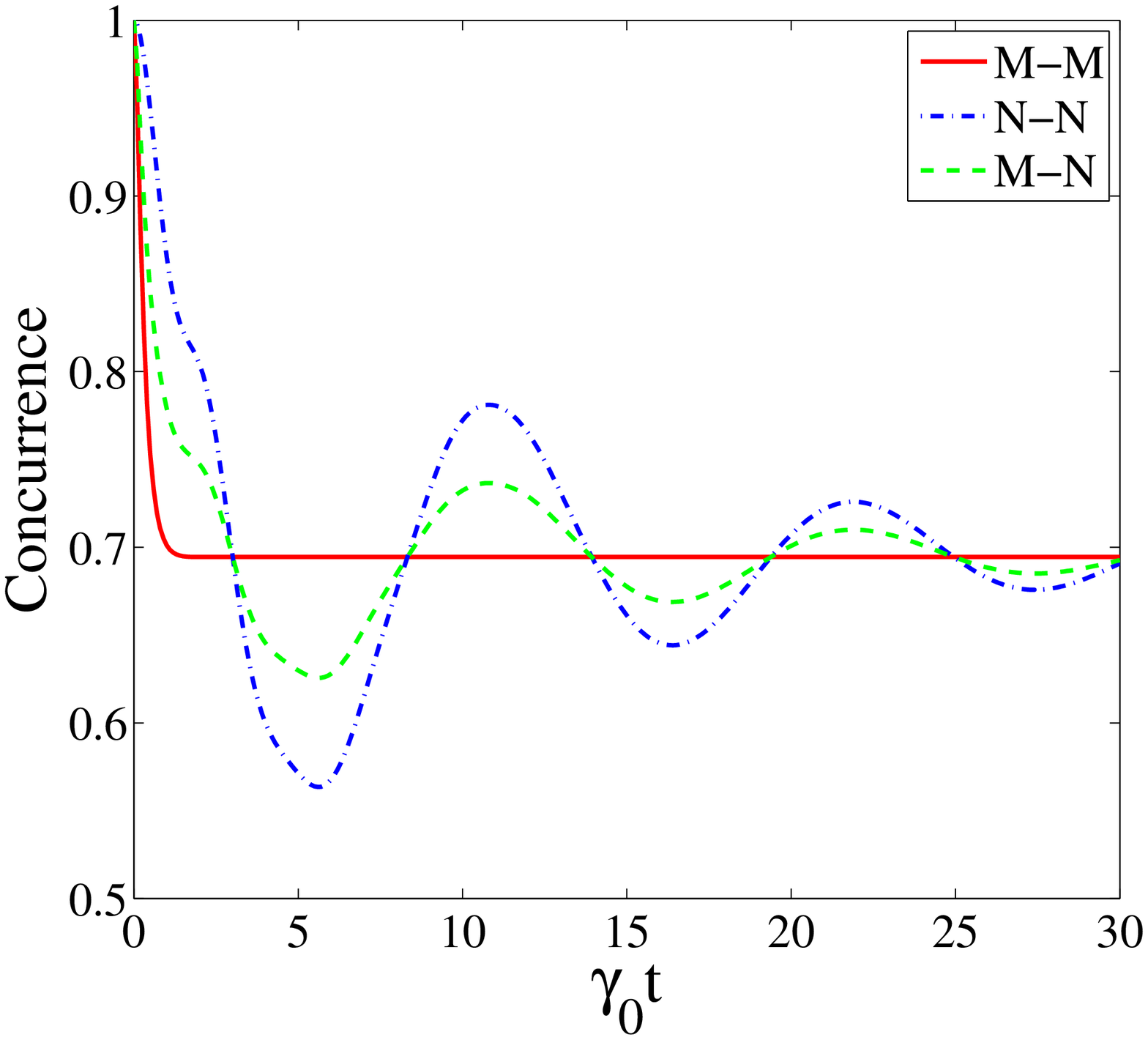}
\caption{} \label{Fig1}
\end{figure}

\newpage
Fig. 6. LBC as a function of $\gamma_{0} t$ with $\Delta_{A}=\Delta_{B}=\Delta_{C}=2$,
(a) Markovian regime with $\lambda_{A}=\lambda_{B}=\lambda_{C}=15 \gamma_{0}$ and (b) non-Markovian regime with $\lambda_{A}=\lambda_{B}=\lambda_{C}=0.5 \gamma_{0}$.
The initial state entanglement is determined by $C_{j}^{A}(0)=C_{l}^{B}(0)=C_{m}^{C}(0)=\frac{1}{\sqrt{3}}$ with $C_{i}^{A}(0)=0$ for $i\neq j$, $C_{k}^{B}(0)=0$ for $k\neq l$, $C_{n}^{C}(0)=0$ for $n\neq m$.

\begin{figure}
        \qquad \qquad\qquad\qquad \qquad a \qquad\qquad \qquad\quad\qquad\qquad\qquad\qquad\quad\qquad\qquad\qquad b\\{
        \includegraphics[width=3.4in]{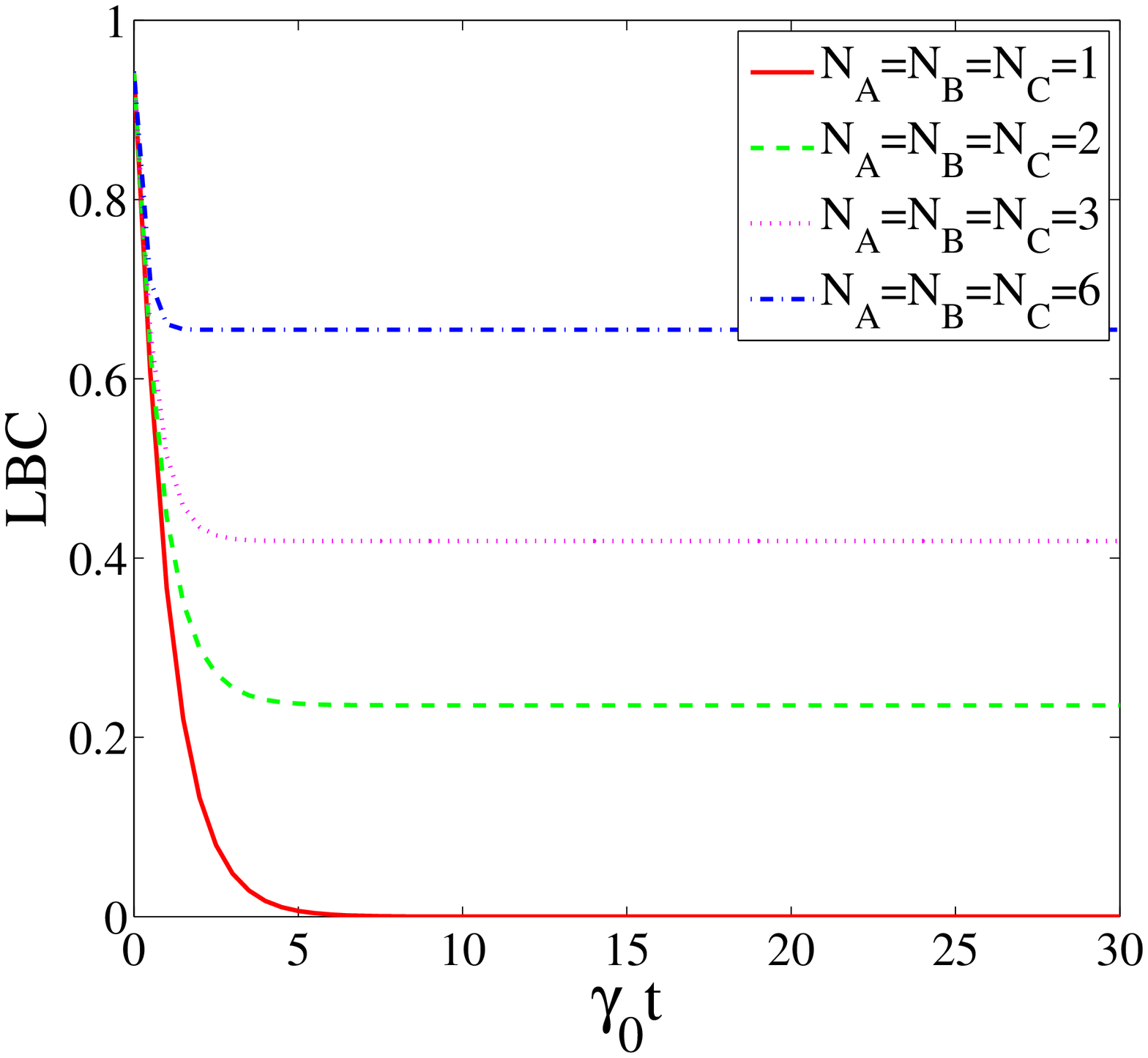}
        \label{fig:first_sub}
    }{
        \includegraphics[width=3.4in]{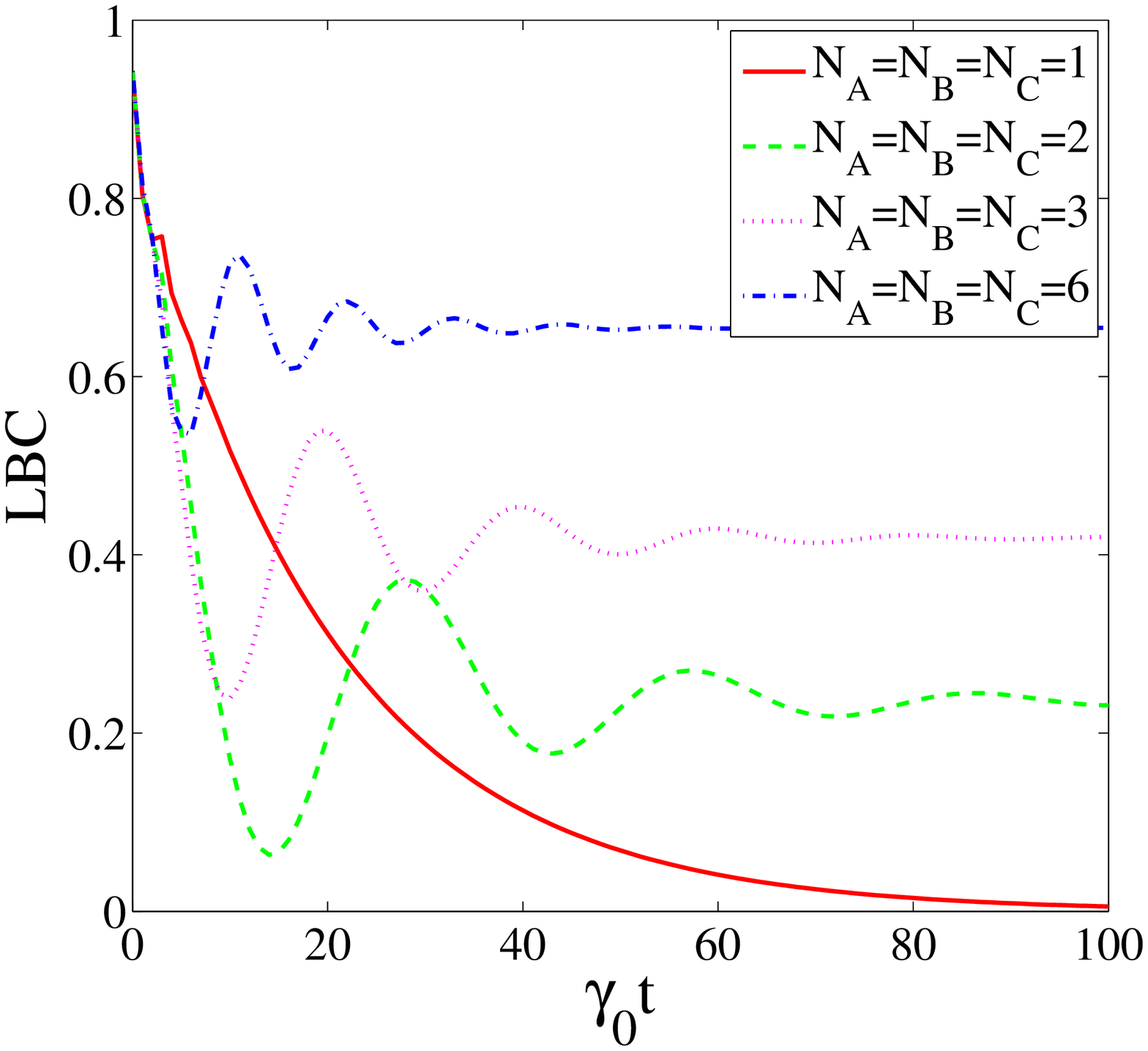}
        \label{fig:second_sub}
    }
    \caption{}
    \end{figure}

\newpage
Fig. 7. LBC as a function of $\gamma_{0} t$ with $N_{A}=N_{B}=N_{C}=6$ and $\Delta_{A}=\Delta_{B}=\Delta_{C}=2$. For the Markovian regime (M), $\lambda_{j}=15 \gamma_{0}$ and for non-Markovian regime (N), $\lambda_{j}=0.5 \gamma_{0}$.

\begin{figure}
\centering
\includegraphics[width=300 pt]{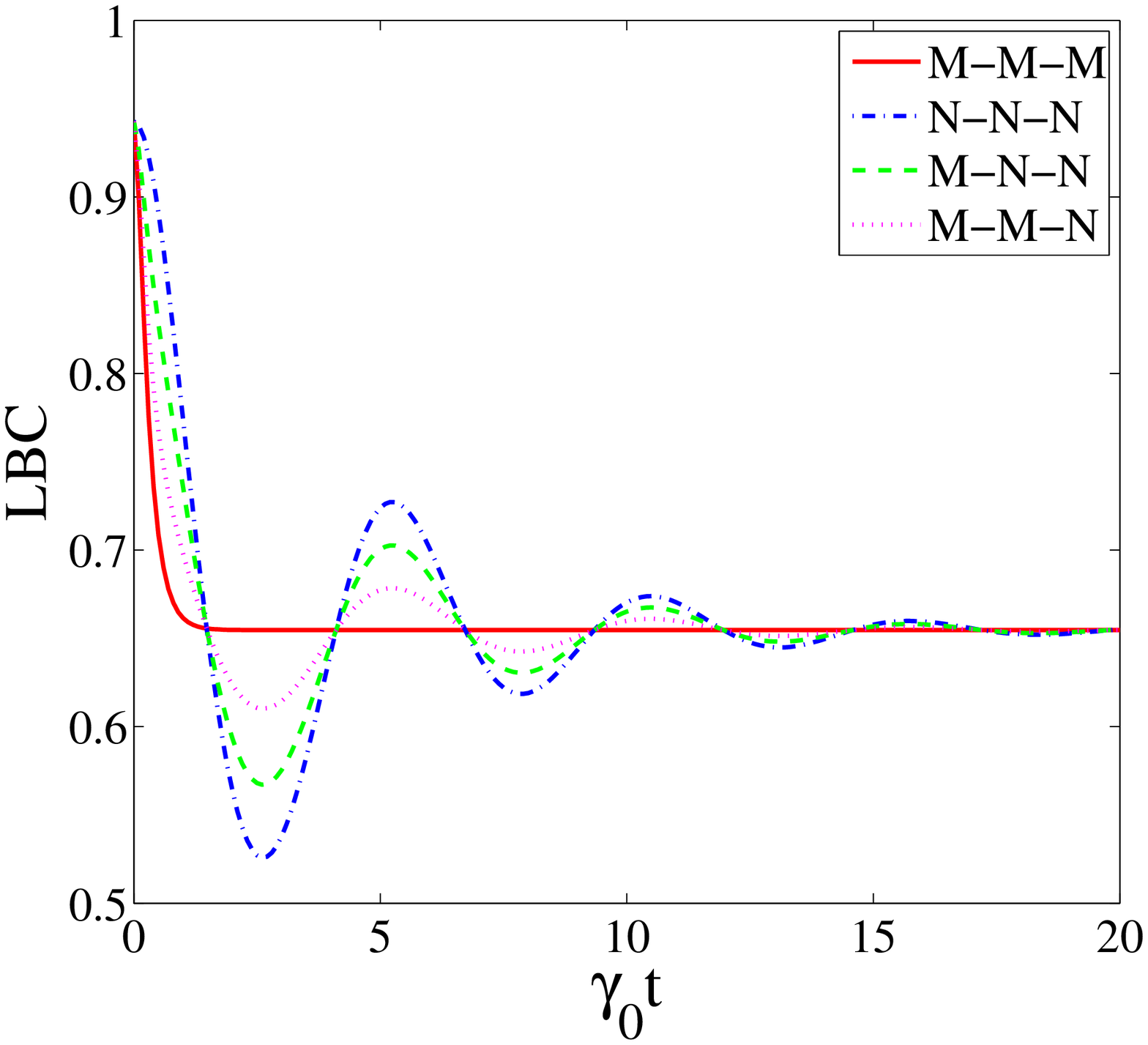}
\caption{} \label{Fig1}
\end{figure}

\end{document}